# A new mapping of technological interdependence

Andrea Fronzetti Colladon [a], Barbara Guardabascio [a], Francesco Venturini [b,c,d,e,*]

[a] University of Perugia, Italy
[b] University of Urbino Carlo Bo, Italy
[c] National Institute of Economic and Social Research, NIESR, UK
[d] The Productivity Institute, TPI, UK
[e] Centre for Innovation Research - Lund University CIRCLE, Sweden

ABSTRACT

How does technological interdependence affect innovation? We address this question by examining the influence of neighbors' innovativeness and the structure of the innovators' network on a sector's capacity to develop new technologies. We study these two dimensions of technological interdependence by applying novel methods of text mining and network analysis to the documents of 6.5 million patents granted by the United States Patent and Trademark Office (USPTO) between 1976 and 2021. We find that, in the long run, the influence of network linkages is as important as that of neighbor innovativeness. In the short run, however, positive shocks to neighbor innovativeness yield relatively rapid effects, while the impact of shocks strengthening network linkages manifests with delay, even though lasts longer. Our analysis also highlights that patent text contains a wealth of information often not captured by traditional innovation metrics, such as patent citations.

## 1. Introduction

Technological interdependence has long been recognized as a driver of innovation and technological change (Rosenberg, 1979). The ability of an innovator to develop new technologies is influenced by the pool of external knowledge available in the economy, which results from prior research successfully conducted by other innovators. In this paper, we look at two sources of technological interdependence (Scherer, 1982a; Archibugi, 1988; Liu and Ma, 2021). The first source is the degree of innovativeness and proximity of neighbors. The closer and more successful neighboring innovators are, the more likely a firm can leverage their knowledge to develop new products or production methods. The second source of technological interdependence is determined by the network of relationships that an innovator maintains within the technology space. A higher number of connections with other innovating entities, as well as a more central position in the innovators' network, increase the likelihood that the innovator can access, assimilate, and integrate external knowledge. This dimension of technological interdependence is reflected in the topology of the innovation network.

The impact of neighbor innovativeness on the success of innovation and on returns to research is a widely discussed topic in the literature (Jaffe, 1989). Technology transfers are channeled by sales of innovative inputs and technology licensing (pecuniary spillovers), and by learning and imitation processes (knowledge spillovers). Input–output analysis has been widely used to gather information on intersectoral technology exchange. This can be observed, for instance, through inter-industry transactions of intermediate or capital inputs (embodied technological change) or through bilateral citation flows among patent documents (disembodied technological change; Keller, 2004). The benefits derived from these factors are directly related to the absorptive capacity of recipient firms (Cohen and Levinthal, 1989) and to the technological proximity of these to other innovators (Jaffe, 1986).

While information on neighbor innovativeness, as a source of technological interdependence across firms and sectors, provides valuable insights into the drivers of innovation, nowadays it may be of limited guidance as modern industrial systems rely on increasingly deeper intersectoral connections (Acemoglu et al., 2016a). The structure of technological linkages, the degree of connectivity among various technology sectors, the position within a densely populated network of innovators all play a significant role in determining the success of innovation and the direction of technological change.

Both the degree of neighbors' innovativeness and the structure (topology) of the innovators' network are recognized as a driver of technological advancement and economic growth, potentially as important as the internal sources of innovation (Romer, 1990; Coe and Helpman, 1995 and Cao and Li, 2019). However, previous research on these two dimensions of technological interdependence has shown minimal overlap. Some studies, for instance, have focused on how technological interdependence induced by knowledge spillovers influences the ability to innovate (Acemoglu et al., 2016b). In contrast, other research has investigated self-generation mechanisms of network linkages and structural interdependence, looking at how existing connections among innovators lead to the formation of additional technology linkages (Taalbi, 2020).

* Correspondence to: Department of Economics, Society and Politics, University of Urbino Carlo Bo, Via Saffi, 42, 61029 Urbino PU, Italy.
*E-mail address:* francesco.venturini@uniurb.it (F. Venturini).

Notwithstanding a long tradition of studies on technology interdependence, there remain several key issues that have not been fully explored in the existing literature. First, it is unclear whether the effect of neighbor innovativeness and that of network structure are related and self-enforcing or, rather, can be seen as complementary dimensions of the same phenomenon. Second, the structure of network linkages that foster technological interdependence among sectors, and affect their ability to innovate, has been scarcely examined. Third, how technology shocks, which alter one of the two dimensions of technological interdependence, propagate from one sector to another and impact their innovation performance remains uncertain.

There are two main motivations behind these gaps in the literature. The first explanation relates to the sources of information used to measure innovation, to capture the characteristics of new technologies, and track linkages among technology sectors. Standard sources include expert surveys, statistics on technology licenses, capital goods purchases, international trade of goods and services, and information from patent documents. Patents are regarded as a highly reliable indicator of new technological ideas introduced to the market (Griliches, 1990). Patents have become increasingly prominent as they provide highly standardized and easily accessible information, which is available on a large scale and covers several aspects of innovation (Hall et al., 2001). The number of patents is often used to quantify the amount of innovative output, whereas prior art claims approximate the breadth of patented innovation. Backward citations accurately describe the derivative nature of innovations, while forward citations capture how current innovations impact the development of future technologies. There are, though, some known pitfalls in using citations (Jaffe and Rassenfosse, 2017). First, there is a marked upward trend in citing, with wide and persistent differences across fields. This raises concerns about the comparability of citations over time and across various technological domains. Second, in several areas, citations are used strategically. Applicants may not disclose prior arts, thus affecting the flow of subsequent citations or, alternatively, may disclose prior arts only where patents are relevant to appropriate returns on their own innovation (e.g., chemicals and drugs) or to block innovation by competitors (e.g., computers and electronics). Third, citation flows are influenced by patent laws and examination procedures. Even within the same jurisdiction, the outcome and timing of patent assessments can vary significantly depending on the examiner (Criscuolo and Verspagen, 2008).

The second (but not less important) explanation of the above-described gaps in the literature relates to how the structure of network linkages is modeled. As earlier mentioned, the most popular approach is to look at direct connections between sourcing (the ‘innovator’) and absorbing entities (the ‘imitator’), assuming that the strength of the linkage is related to their technological proximity (Jaffe, 1986). Extensions of this approach include country-level analyses based on measures of trade and geographical distance (Coe and Helpman, 1995; Madsen, 2007) and sectoral-level analyses based on within- and cross-country measures of intermediate input transactions (Scherer, 1982b; Verspagen, 1997b; Pieri et al., 2018). Only a limited number of studies have looked at how indirect linkages model technology interdependence across sectors. Leoncini et al. (1996) use network analysis on input–output relations to track international differences in the technological system: for example, Germany exhibits dense and evenly distributed intersectoral innovation linkages, whereas Italy has a limited number of high-tech sectors coexisting with a large pool of traditional sectors. Acemoglu et al. (2016b) use citation networks to map the linkages across US technology fields and predict their innovation capacity. Cao and Li (2019) measure technology applicability using citation networks and predict the contribution of each sector (node) to knowledge development within the entire technology space (network).[1] Taalbi (2020) examines which factors affect the creation of new inter-industry technology linkages, proxied by innovation commercialization between sectors, finding that direct connections have a greater impact than indirect ties.

Drawing upon the foundations of existing literature, this study examines how interdependence in the technology space influences the creation of new knowledge. Specifically, we explore how the innovativeness of neighboring entities and the structure (topology) of network linkages, which reflects the position of the sector within the technology space, collectively contribute to the success of innovation activities. Through text mining techniques, we analyze the abstract of 6.5 million patents granted by the United States Patent and Trademark Office (USPTO) between 1976 and 2021 and apply network analysis to uncover the strength of technological interdependence among different sectors (classes). We design a knowledge production function in which innovation output depends, along with standard determinants, on the degree of neighbor innovativeness and the strength of structural linkages within the innovation network. Using data related to 128 technology sectors, we first estimate our empirical model with panel dynamic regression methods and then assess the response of innovation output to technology shocks affecting both dimensions of technological interdependence.

Our first key finding is that both neighbor innovativeness and the network structure play a crucial role in shaping sector innovation performance. This finding is significant as it bridges two streams of empirical evidence that have previously developed with minimal overlap. The effect of these two factors is quantitatively comparable in the long run. As a second key piece of evidence, our research demonstrates that a shock increasing the degree of neighbor innovativeness leads to a higher level of sector innovation within a relatively short time frame (approximately in less than five years) but then vanishes out. Conversely, the effect of positive shocks affecting the network of structural linkages takes longer to become statistically significant and economically impactful (more than five years); however, the latter effect is more long-lasting. In general, the responsiveness of sector innovation to unanticipated changes in both dimensions of technological interdependence has arisen only in the most recent decades. Our third main result highlights the significant impact of structural linkages through various measures of network centrality, including degree centrality, betweenness, closeness, and distinctiveness (Freeman et al., 2002; Fronzetti Colladon and Naldi, 2020), as well as the Katz metric of centrality (Katz, 1953). The paper exploits information conveyed by the full set of network centrality measures to construct a multi-dimensional (latent) factor that summarizes variation in innovation network linkages that is relevant to predict sectoral innovation (see Lanjouw and Schankerman, 2004). We demonstrate that the latent factor possesses a greater explanatory power for assessing the impact of structural linkages compared to any individual measure of network centrality from which it is derived. Finally, as a further novel piece of evidence we show that, when using patent text information, the impact of technological interdependence (broadly intended) is larger than when using traditional (citation-based) measures. Patent texts may indeed capture intersectoral linkages over a broader set of technical features, maybe induced by incremental innovations, whereas patent citations are likely to trace intersectoral linkages related to (parts of) leading technologies. We document the robustness of all these results in many dimensions, namely (i) the measure of patent output (simple vs. quality-adjusted patent counts), (ii) the proxy for technological distance (random vs. text similarity), (iii) the modeling of unobservable factors (time dummies vs. common correlated effects), (iv) cross-sector differences (slope homogeneity vs. heterogeneity), and finally, (v) the estimation procedure (log-linear vs. count data regression).

[1] Hotte (2023) uses a two-layer, two-way related network to study the impact of inter-industry interactions on various dimensions of US industry performance. The upper layer is based on input–output (trade) relations, and the bottom layer on inter-industry technology (citation) relations. The latter linkages appear to dominate as showing positive effects both horizontally and vertically.

This paper addresses three main strands of the literature. First, we contribute to a deeper understanding of the inter-sectoral influences in innovation processes (Castellacci, 2008), sectoral patterns of innovation and technological specialization (Pavitt, 1984; Malerba, 2002; Archibugi et al., 2023), as well as the trajectories of technological development (Dosi, 1982). In this regard, we illustrate the growing importance of technological interdependence for innovation development and the increasing complexity of this interdependence over time. Second, our research contributes to the debate surrounding the puzzling decline in research productivity and its correlation with the (potential) decrease in knowledge spillovers. This phenomenon may be prompting companies to intensify their own R&D efforts in order to maintain consistent rates of innovation (Venturini, 2012; Bloom et al., 2020). Our research indicates that technological interdependence has significantly grown in recent years. This would exclude any causal nexus between changes in the effects of knowledge spillovers and structural linkages (which are increasing) and changes in returns to research (which are decreasing). Third, we contribute to improving the measurement of technological change, showing that new data methods are a powerful tool to study innovation at a granular level and, in the meantime, identify aggregate technological trends (Scherer, 1983; Kelly et al., 2021). More in detail, we show that technological interdependence can be well gauged through textual analysis of patent documents, in addition to more traditional measurement approaches.

The remainder of the article is organized as follows. Section 2 briefly reviews the literature. Section 3 presents the empirical model. Section 4 describes data sources and provides summary statistics. Econometric results are reported in Section 5, while Section 6 outlines the conclusions of our study.

## 2. Literature review

Our research converges at the nexus of various strands of the literature. These include studies on knowledge spillovers and technology complementarities, the evolution of technological systems, and, not less importantly, new data methods for innovation measurement.

### 2.1. Innovation and technological interdependence

Innovation is an original piece of knowledge that expands the existing state of technological knowledge. Innovation is created in response to changes in firms' demand conditions, in the opportunities offered by technology pushes (Mowery and Rosenberg, 1979), and is developed by exploiting both internal or external knowledge sources (Pavitt, 1984), following sectoral technology patterns (Malerba and Orsenigo, 1997). Among external sources of innovation, a major attention of the literature has been paid to knowledge spillovers and the conduits of technology flows between source and recipient entities of technological knowledge (Schmookler, 1966; Scherer, 1982b,a; Verspagen, 1997a). Most works build inter-industry matrices of technology transfer by extrapolating information from thematic surveys, technology licenses, or looking at patent citation flows (Archibugi and Pianta, 1996). All these measures capture transfers of disembodied technological knowledge, meaning the knowledge is not incorporated in any input exchanged between firms or sectors. Other important pieces of knowledge are embodied, and spread through the exchange of intermediate inputs and high-tech investment goods, as well as through the movement of workers between different jobs (Keller, 2004; Mendi, 2007; Venturini, 2015).[2] However, knowledge spillovers are limited in their geographical scope and tend to be spatially concentrated. Peri (2005) documents for the US that "only" one-fifth of knowledge is exploited outside the geographical area of creation. Bottazzi and Peri (2003) find that knowledge spillovers in Europe remain confined within a 300 km radius from the place of innovation.

A related line of studies examines the vertical transmission of innovation shocks, i.e., how innovation in upstream technology sectors transmits downstream, favoring innovation of technology users. Tracking vertical linkages through the citation network, Acemoglu et al. (2016b) document that upstream innovation explains 14% of panel variation in innovation achievements of downstream technology sectors. Funk and Owen-Smith (2017) gauge how new technologies are used by later technologies or alter the use of earlier technologies through network-based metrics, categorizing innovations either as consolidating or destabilizing. A common finding in this literature is that the structure of linkages, sometimes modeled as networks, have a stable architecture made by key hubs linked to numerous downstream users, and that these connections affect the creation of new ties and the development of further innovation.

Evolutionary studies of innovation examine the systemic mechanisms that lead to novelty and the development of breakthrough technologies. Innovation is seen as an original recombination of existing pieces of knowledge developed in different branches of the economy (Weitzman, 1998). A specialized knowledge base is considered a key requisite to innovate. Nonetheless, knowledge diversification offers relevant gains in the development of new technologies, due to knowledge cross-fertilization and recombination and the diversification of innovation risk (Garcia-Vega, 2006). The degree of knowledge relatedness determines whether knowledge created in one sector is easily exploitable in other sectors or geographical areas (Frenken et al., 2007; Castaldi et al., 2015). Related knowledge variety is, on average, positively correlated with the rate of innovation. Conversely, breakthrough innovations originate from combining unrelated variety knowledge, open up new domains for technological advancement and pave the way for additional incremental innovations by recombining related knowledge varieties (Schoenmakers and Duysters, 2010).

Innovation can also be viewed as a development of "adjacent possibles" (Kauffman, 2000). Novelties emerge from the interaction among different forces in complex systems (natural, socio-economic, technological), result from the combination of past discoveries, and develop in areas once thought difficult to reach. The emergence and distribution of novelties follow well-defined statistical laws, such as the Heaps' law and Zipf's law, respectively (Tria et al., 2014). Novelties emerge in every field of human activity. Initially, they coexist and compete with older and concurrent ideas. Over time, they rapidly gain popularity through a self-reinforcing mechanism, often described as "the rich get richer" mechanism, ultimately prevailing over other ideas (Monechi et al., 2017). In this context, Taalbi (2023) shows that the structure of the product (or technology) space is a good predictor for the new areas in which firms will innovate: the firm development of new product types (commercialized innovations) depends on search scope (the firm's share of cited patent classes) and search depth (the firm's proportion of cited patent classes relative to the recent past). In a related paper, Taalbi (2020) studies the evolution of the technological system in Sweden, using innovations commercialized to other sectors as a proxy for new technological ties: 30% of variation in these new connections can be attributed to pre-existing network linkages. Similarly, Kim and

[2] Hanley (2017) studies within-industry dependence in innovation processes (so-called innovation sequentiality) by looking at patent transfers between companies active in the same technological field: industries with greater sequentiality are found with higher rates of innovation and profitability.

Magee (2017) use patent citation flows to predict changes in the topology of innovation networks across US technology sectors.

One topic that has been extensively explored in the literature is the diversification of a firm's patent portfolio. These strategies follow trajectories reflecting the ties and distance among the technological fields in which innovating companies are active (Breschi et al., 2003). Companies often engage in technological diversification before expanding their offerings, as the development of new products involves leveraging a diverse range of technologies (Pavitt, 1998). Historically, there has been a notable alignment between technological and productive activities in which firms engage (Teece et al., 1994). Market and technological diversification are driven by knowledge coherence, as firms gradually shift towards areas of the product and the technology space where the knowledge required is close to their competencies. However, according to recent evidence, product diversification would anticipate technological diversification (Piscitello, 2000). Furthermore, for the majority of firms, the extent of product diversification would be greater than that of technological diversification (Dosi et al., 2017).

### 2.2. New data methods for innovation measurement

In the literature on innovation and technological change, patent documents have long been used as a primary source of information. At the firm level, patents are found to be significantly related to various dimensions of performance, such as productivity or market value (Hall et al., 2005). At the aggregate level, the nexus between patenting and productivity performance has been less clear, probably due to mismeasurement issues, and the effect of confounding factors such as institutions, etc. (see Nagaoka et al., 2010). However, Madsen (2008) and Kogan et al. (2017) have recently shown that the rate of patenting and breakthrough innovations are positively related to the growth rate of GDP per capita (or per worker) in the very long run.

In recent years, significant advancements have been made in collecting information from patent documents, largely due to the implementation of advanced data analysis techniques (Arts et al., 2021). Machine learning-based textual analysis has transformed research in this field by overcoming numerous limitations of traditional measures of patented innovations. Patent text is characterized by precise technical language and high word standardization, allowing for a highly accurate assessment of innovation. Semantic patent text search enables a more seamless analysis of innovation with respect to the use of patent metadata, which are crystallized along well-defined (pre-packaged) criteria (citations, claims, etc.). Text extraction from patent documents is fruitful for inferring the technological proximity between innovating firms and the level of technological interdependence existing across sectors, while business documents provide processable information on companies' innovation strategy (Fattori et al., 2003).[3]

Bergeaud et al. (2017) is one of the first works using patent text information to study technological development. These authors devise a categorization based on the content of the USPTO patent abstracts and compare these groupings with technological (IPC) classes along various dimensions (diversity, originality, generality, etc.). For instance, the citation rate of patents belonging to the same semantic class is significantly higher than that of patents within the same technological class. This finding suggests that traditional systems of classification could produce imperfect categorization of innovations (see Moed et al., 2006; Lafond and Kim, 2019).

Arts et al. (2018) build text-based similarity indicators for the entire corpus of USPTO patents. These measures are able to reproduce earlier findings of the literature on localized knowledge spillovers based on standard citation metrics but present a much higher statistical reliability. Measures of patent text similarity (cosine proximity) reveal, for the US, a local concentration of knowledge spillovers weaker than that emerging from citation flows (Feng, 2020). As discussed above, this may reflect the firm's strategic use of patents, the fact that citation paths are influenced by the background of examiners or attorneys and, not less important, that patent documents convey a larger body of information about the innovation than citation streams.

Gerken and Moehrle (2012) develop an index of innovation novelty constructed by comparing the semantic structure of patent documents filed at distant points in time. Kelly et al. (2021) gauge innovation radicalness (significance) with the ratio of patent text measures of forward and backward similarities, finding that groundbreaking innovations drive long-term economic growth in the US. Carvalho et al. (2021) investigate innovation strategies of US firms active in new technological areas by mining their patent documents, detecting a positive association between the strategy of innovation exploitation and sales growth. Mann and Püttmann (2023) use keyword search analysis on US patent documents to measure automation innovations, mapping sectors in which these technologies are developed and sectors in which they are used. Park et al. (2023) use network analysis to build similarity measures for a collection of patents and scientific publications using information on citations and document texts. The results of their research suggest that new technologies are currently less disruptive than in previous eras, indicating a potential slowdown in research productivity and in the rate of technological progress.

## 3. Empirical model

Our analysis provides new insights into how technological interdependence impacts sectoral innovation, considering the role of neighbor innovativeness and related knowledge spillovers, as well as the structure (topology) of the innovation network and the position that each sector holds within the technology space.

We assess how technological interdependence affects innovation by estimating a knowledge production function at the level of technological sectors (patent classes). We assume that new knowledge ($\Delta N$) is created thanks to the absorption of knowledge developed by other sectors (neighbor innovativeness) and the linkages that each sector maintains with other innovative entities (network structure). Formally, $\Delta N$ is assumed to depend on $L$, which is our proxy for technological interdependence (broadly intended) among technology areas, while $\alpha$ identifies the effect of this force on the capacity of each sector to create new knowledge (innovation):

$$\Delta N = f(L) = L^{\alpha}. \quad (1)$$

Eq. (1) can be extended to include other standard drivers of knowledge generation, such as the cumulative value (stock) of knowledge developed within each technology area, $N$. $N$ reveals whether the state of technological knowledge generated in the past affects the output of current innovation processes, by favoring the intertemporal (within-industry) transmission of knowledge (*standing-on-the-giants'-shoulders* vs *fishing-out* effects; Caballero and Jaffe, 1993). Following Ha and Howitt (2007), we extend Eq. (1) in two further respects. First, we account for the effect of the current innovation effort, $R$ (R&D input), that could reinforce the intertemporal transmission of knowledge. Second, we consider the degree of product diversification of the sector, $Z$, that may (fully or partially) outweigh the stimulating effect of the other two internal drivers of innovation ($R$ and $N$):

$$\Delta N = f(L, N, R, Z) = L^{\alpha} \cdot N^{\beta} \cdot R^{\gamma} \cdot Z^{\delta}. \quad (2)$$

We estimate the stochastic, log-linear version of Eq. (2) that considers two sources of intersectoral technology dependence, namely the degree of neighbor innovativeness and the intensity of the structural linkages within the innovation network, respectively denoted as

[3] See Nathan and Rosso (2022, 2015) for a study on the mapping of digital firms and their innovation performance based on text mining of data collected through the scrapping of company websites.

$L^{NI}$ and $L^{NS}$. Below, Section 4 describes how these two sources of technological interdependence are measured.

$$\ln \Delta N_{il} = a_i + \underbrace{\alpha^{NI} \ln L_{it}^{NI}}_{Neighbor\ Innovativeness} + \underbrace{\alpha^{NS} \ln L_{it}^{NS}}_{Network\ Structure} + \beta \ln N_{it} + \gamma \ln R_{it} + \delta \ln Z_{it} + \epsilon_{it} \quad (3)$$

In Eq. (3), $\Delta N$ is defined as the flow of new patents granted to each sector $i$ at any point in time $t$. The impact of $L^{\nu}$ (with $\nu = NI$ or $NS$) may be positive when the success of innovation activities is self-sustaining across sectors due to knowledge spillovers, technological complementarities, etc. ($\alpha^{\nu} > 0$), or negative because of research cost inflation (resource extraction), innovation duplication or lock-in effects ($\alpha^{\nu} < 0$). $N$ should capture dynamic (intertemporal) returns to innovation ($\beta > 0$): these could be highly persistent because of the cumulative effects of knowledge creation over time ($\beta \sim 1$) or diminish due to the exhaustion of technological opportunities ($\beta < 1$). $R$ reflects the size of purposeful innovation (R&D) effort, which is undertaken to expand the state of technological knowledge ($\gamma > 0$). $R$ is usually defined in terms of human resources allocated to innovation processes. $Z$ reflects the number of companies (applicants) engaged in innovation activities in each sector (class). A negative value for the coefficient of this variable would indicate that innovation effort increases less than proportionally with the number of innovators, thus reducing aggregate returns to R&D ($\delta < 0$). By contrast, a positive value would indicate that companies have the opportunity to leverage economies of scope and achieve greater returns in sectors with a higher concentration of innovative firms ($\delta > 0$). The effect of the systematic (time-invariant) differences existing across sectors in the patent propensity, to engage in innovation networks, etc., is captured by sector-specific fixed effects ($a_i$).

In our regression model, the effect of common exogenous shocks is primarily accounted for by expressing all variables in deviation from the yearly (cross-sectional) average. This is equivalent to using common time dummies and is helpful to neutralize the bias associated with weak levels of residuals' cross-sectional correlation in innovation processes (Cross-Sectional Dependence, CSD). However, in robustness checks, we control for strong cross-sectional dependence by including common correlate effects (CCE) in the specification. These terms are calculated as cross-sectional averages of all (not demeaned) variables in the model, and serve to mitigate the bias associated with co-movements induced by "third factors" such as technology, trade or demand shocks, having a differentiated impact withinin the technology space (i.e. across sectors). Eberhardt et al. (2013) emphasize the importance of properly controlling for the impact of unobservable common factors. Failure to account for these effects can lead to misinterpret them as evidence of knowledge spillovers, as the latter are typically measured using the proximity-weighted average of innovation efforts (or outcomes) of neighboring entities, such as firms, industries, regions, or countries.

It should be stressed that Eq. (3) models the long-run (equilibrium) relation of technological interdependence existing among sectors. However, in light of the long-time dimension of our data (see below for details), the regression is estimated with a dynamic specification, i.e., as an autoregressive distributed lag model (ARDL). This procedure ensures consistency of long-run estimates irrespective of the integration order of the variables, and is robust to reverse causality when the lag structure is optimally specified. Below, we report the long-run parameters estimated for Eq. (3). These can be interpreted as elasticities.[4]

[4] Considering a general long-term relation of the following type, $y_{it} = a_i + bx_{it} + \epsilon_{it}$, the corresponding dynamic specification with one-year lag of the variables is formulated as $y_{it} = a_i + a_1 y_{it-1} + a_2 x_{it} + a_3 x_{it-1} + \epsilon_{it}$. From the latter, one can then recover the long-term effect of the explanatory variable as $b = (a_2 + a_3)/(1 - a_1)$.

## 4. Data sources, methods and variable description

### 4.1. Patent data and text mining

We perform our analysis on the universe of utility patents granted by the US Patent and Trademark Office (USPTO) between 1976 and 2021. The USPTO patent data is a valuable source as it offers a comprehensive overview of the most important world's technology market, providing highly reliable information on several characteristics of patented inventions. This data has been utilized to gain insights into technology trends, analyze firm performance, and inform strategic decision-making in various industries (Griliches, 1990; Hall et al., 2001; Hall and Harhoff, 2012). The majority of papers in the literature have used coded information on names and locations of applicants (or inventors), and on the characteristics of their inventions (such as cites made and received, technological classes, and co-patenting processes). However, the USPTO now provides the entire bulk of patent documents in a machine-readable format, enabling the mining of these texts and the creation of more sophisticated measures of innovation content.[5]

We analyze the abstracts of 6,497,894 patents for which we have relevant information on application date, technological class, etc. This approach offers the benefit of concentrating on concise texts that have remained largely consistent over a period of half a century. Patent abstracts have not been significantly affected by changes in patent laws that modified the requisites of patentability, the examination process and, as a consequence, the timing and quality of these procedures. van Pottelsberghe de la Potterie (2011) outlines the evolution of the US patent jurisdiction since 1980, highlighting the impact of various Supreme Court decisions. These have gradually expanded the scope of patentable subjects (including genetically engineered bacteria, software, business methods, financial service products, and more), while also relaxing the novelty requirements for obtaining a patent. All this significantly increased application workload, which lowered the quality of the examination process and, in turn, stimulated the demand for patent protection for low quality innovations (see also Jaffe, 2000).

We implement our study by assigning patents to 3-digit technological categories resulting from the Cooperative Patent Classification (CPC) and classifying them according to their application date. We examine patent abstracts using the SBS BI software, which allows to conduct advanced textual analyses and create semantic networks (Fronzetti Colladon and Grippa, 2020). The procedure has been implemented with the following steps. First, we preliminary remove punctuation, stop-words, and special characters (Perkins, 2014) then, after lowercasing the text, we extract the stems through the Porter algorithm (Willett, 2006).[6] Second, we assemble the vectors of abstracts into a corpus-term matrix with sector/year by rows and word occurrences by column. Cells in the matrix assume the value of 1 if the column term appears at least once in a specific set of abstracts (row), and 0 otherwise.[7] Third, we exclude highly common terms that appear in more than 75% of abstracts, and rare terms that appear in less than 0.1% of these documents. The analysis is conducted on the most recurrent terms in the resulting vocabulary (up to a maximum of 15,000 words). Subsequently, we apply the well-known Term Frequency Inverse Document Frequency (TFIDF) transformation to the corpus-term matrix (Roelleke and Wang, 2008). This transformation assigns greater importance to the most recurrent terms in patent documents but that, at the same time, are not common across all technology sectors. Lastly, we use the L2 normalization to account for differences in the number and length of abstracts across technology sectors. In practice, we rescale the row vectors so that the square of their cells sums up to one, $V_{it} = TFIDF_{it}/\|TFIDF_{it}\|$ (see Kelly et al., 2021). This matrix serves for the construction of the similarity network which, as described in the next section, exploits information on cosine similarity between the cells of the matrix rows.

[5] Patent text data are retrieved from the following link: https:/patentsview.org/download/detail_desc_text.

[6] For example, the terms "beauty" and "beauties" would both be transformed into the word "beauti".

[7] As discussed later, we also exploit an alternative approach by populating the matrix cells with word frequencies found within patent abstracts. However, this method does not yield significantly different results. This suggests that assessing the presence of a term within a set of patents for one technology sector in one year may be sufficient to determine its similarity to the other sectors.

To illustrate the outcome of our text mining process, we consider the abstract of three hypothetical patents as an example. Abstract 1: *This invention discloses a machine-learning model for predicting the maintenance needs of industrial machinery. The model utilizes sensor data and historical maintenance records to identify patterns and predict potential failures before they occur.* Abstract 2: *This patent concerns an AI-powered system for proactive maintenance of industrial equipment. The system leverages sensor data analytics and machine learning algorithms to anticipate equipment failures and optimize maintenance schedules.* Abstract 3: *This invention concerns a chemical composition for improving the adhesive properties of a bonding agent. The composition comprises a unique blend of polymers and additives that enhance the strength and durability of the bond.* The first two patents are clearly more similar to each other than the third one, as both refer to "machine learning" and "maintenance", which are terms that appear in their abstracts but are less frequent in the overall corpus of patent documents (common words such as "the", "an", and "of" are filtered out during the pre-processing).

### 4.2. Measuring technological interdependence

We measure technological interdependence by constructing proxies for the degree of knowledge spillovers (neighbor innovativeness), and for the topology of structural linkages and the sector position within the technology space (network structure). To construct these measures, we primarily exploit information extracted from the text of patent abstracts. However, to gain insights into the informative advantage offered by this methodology, we also construct measures of sectoral interdependence based on bilateral citation flows, following the main practice in the literature. When analyzing network structure (topology), we take into account the influence of both direct linkages and indirect connections by employing several measures of sector (node) centrality within the technology space (network).

*Neighbor innovativeness*

Technological interdependence between sectors $i$ and $j$, fueled by neighbor innovativeness, is measured with the proximity-weighted mean of their innovation (patenting) capacity (that is, our outcome variable $\Delta N$), for any year between 1976 and 2021:

$$L_i^{NI} = \sum_{j=1}^{n} w_{ij} \Delta N_j \quad with \quad w_{ij} = 0 \quad if \quad i = j \tag{4}$$

The subscript $t$ is omitted wherever possible for sake of brevity. Proximity weights derived from patent texts are computed using a cosine similarity index, defined as $w_{ij} = \rho_{ij} = V_i \cdot V_j$ with $\rho_{ij} \in [0,1]$, where $V$ is the corpus-term matrix described in the previous subsection. By contrast, proximity weights based on patent cites are computed using the relative flows of bilateral citations, defined as $w_{ij} = c_{ij}/\sum_j^n c_j$ where $c_{ij}$ identifies the number of patent citations made by sector $i$ to patents of sector $j$ over the total number of citations made by the former sector. Below, we denote as $\mathbf{W}$ the matrix of weights, with $ij$-element defined by $w_{ij}$.

Our metric of neighbor innovativeness builds upon Acemoglu et al. (2016b). One key difference between these two measures lies in the fact that cosine similarity does not reveal the underlying origin of the ideas that are underneath innovations. Instead, it measures the proximity between earlier and subsequent innovations based on their descriptions in the patent abstract. In contrast, patent citations trace the vertical (unidirectional) transmission of technological knowledge between cited and citing innovations.

*Network structure*

The structure (topology) of connections among innovators is another important factor driving the success of the sector's innovation activities. Differently from the other dimension of technological interdependence, which relies upon innovation capacity and proximity of linked sectors, the structure of the innovation network reflects the position of each sector in the technology space and the intensity of inter-sectoral linkages. In the network, each node corresponds to a technology (innovating) sector, and the arcs connecting the nodes reflect the relationships existing among sectors (innovators). The intensity of the linkages is gauged by weighting the arcs between nodes with the pairwise similarity scores (constructed as detailed above).

Our metrics of structural linkages come from social network analysis (Wasserman and Faust, 1994). We measure the network centrality of each technology sector using well-known metrics earlier used in the analysis of patent citations, such as degree, betweenness and closeness centrality (Katz, 1953; Hung and Wang, 2010; Liu et al., 2021; Sternitzke et al., 2008). Furthermore, we also take into account the newly developed metric of distinctiveness centrality, which offers the advantage of capturing exclusive connections between technology sectors (Fronzetti Colladon and Naldi, 2020). We produce a similarity network for each year of the time interval of our study. By construction, these networks are complete and symmetrical. However, in order to streamline their structure, we eliminate arcs with minimal similarity scores, specifically those that fall within the lowest quartile of the similarity distribution. The set of centrality measures adopted allow to consider both direct and indirect linkages within the network.

The first metric utilized is the centrality index developed by Katz (1953). This measure sums up the number of arcs $l$ (linkages) existing between nodes, and weighs these connections by a decay (penalty) parameter $\eta$ (with $\eta \in (0,1)$) that penalizes paths in relation to their length: when $\eta$ is low (high) a greater (lower) weight is attached to the shortest path length (Liben-Nowell and Kleinberg, 2003; Taalbi, 2020):

$$L_i^{NS,KZ} = \sum_{l=1}^{\infty}\sum_{j=1}^{n} \eta^l w_{ji}^l = 1 + \eta \sum_{j=1}^{n} w_{ji} + (\eta \sum_{j=1}^{n} w_{ji})^2 + .. + (\eta \sum_{j=1}^{n} w_{ji})^{\infty} \tag{5}$$
$$\vec{L}^{NS,KZ} = ((\mathbf{I} - \eta \mathbf{W}^T)^{-1} - \mathbf{I})\iota.$$

The Katz index, $\vec{L}$, is defined as a vector (denoted by $\rightarrow$) in which each element $i$ is taken in absolute terms. $\mathbf{W}$ is the similarity matrix based on patent text (or bilateral citations described above), and the subscript $T$ denotes its transpose. $\mathbf{I}$ is the identity matrix, whilst $\iota$ is a vector of size $n$ consisting of ones. In the first formulation of Eq. (5), the convergence of summation is ensured if $1/\eta$ is larger than the greatest eigenvalue of $\mathbf{W}$. Since each cell in $\mathbf{W}$ is a measure of the direct linkage between each pair of sectors, a valuable property of the Katz measure of centrality is that the overall (structural) linkages can be decomposed into the sum of direct (first-order) linkages ($L^{NS_D,KZ}$) and indirect (higher-order) linkages ($L^{NS_I,KZ}$):

$$\vec{L}^{NS_D,KZ} = \iota\mathbf{W} \qquad \vec{L}^{NS_I,KZ} = \vec{L}^{NS,KZ} - \eta(\iota\mathbf{W}). \tag{6}$$

In the two equations above, the superscript $NS$ stands for Network Structure, $KZ$ for Katz, whilst the subscripts $D$ and $I$ for Direct and Indirect linkages.

Another key measure that we use to capture direct linkages in the technology space is the index of *Degree centrality* (Wasserman and Faust, 1994). This measure reflects the number of connections that a node has within a network. In networks where connections have a direction, it is possible to differentiate between incoming and outgoing arcs. The total number of incoming arcs is referred to as in-degree, while the number of outgoing arcs is known as out-degree. For each node (sector) $i$, the Degree centrality formula used is:

$$L_i^{NS_D,DG} = DG(i) = \sum_{j=1,j\neq i}^{n} I(w_{ij} > 0) \tag{7}$$

where $I(w_{ij} > 0)$ is a function that assumes the value of one if there is an arc connecting nodes $i$ and $j$ with a positive weight, and zero otherwise. In order to ensure comparability across networks of varying sizes, we standardize degree centrality by dividing the index by $(n-1)$. In the weighted version, degree centrality is calculated by adding up the weights of the arcs connected to a node. For instance, if patents from sector A are cited 100 times in total by three other sectors, the in-degree of A would be 3, as the sector has three incoming connections; however, the weighted in-degree would be 100, considering the total weight of incoming arcs. It is also worth noting that, for the purpose of our analyses, we exclude self-loops. In Eq. (7), the superscript $NS_D, DG$ stands for Network Structure measure of Direct linkages based on DeGree centrality.

To gauge the effect of indirect connections, we consider an additional set of network centrality measures, namely the indexes of (i) betweenness, (ii) closeness, and (iii) distinctiveness centrality. *Betweenness centrality* quantifies how often a node lies in the shortest path connecting each pair of other nodes, reflecting thus its brokerage power (Wasserman and Faust, 1994). The weighted version of this index, which is useful in our case as similarity and citation networks are particularly dense, is obtained considering the inverse of arc weights to calculate network distances (Opsahl et al., 2010). This means that arcs with a higher number of citations, or a greater text similarity, will be considered closer when calculating the shortest paths. For node $i$, the Betweenness index is calculated as:

$$L_i^{NS_I,BE} = B(i) = \sum_{j<k} \frac{d_{jk}(i)}{d_{jk}} \tag{8}$$

where $i$ is distinct from nodes $j$ and $k$. $d_{jk}$ is the total number of the shortest paths connecting nodes $j$ and $k$, and $d_{jk}(i)$ is the number of paths that include node $i$. The index is divided by $(n-1)(n-2)/2$ for undirected graphs, and by $(n-1)(n-2)$ for directed graphs, to make it comparable across networks of different sizes. In Eq. (8), the superscript $NS_I, BE$ indicates Network Structure measure of Indirect linkages based on BEtweenness centrality.

*Closeness centrality* determines the proximity of a node to all other nodes in a network. It is calculated as the inverse of the sum of the shortest path lengths from that node to all the others (Wasserman and Faust, 1994). For node $i$, the Closeness index is given by

$$L_i^{NS_I,CL} = C(i) = \frac{n-1}{\sum_j d(j,i)} \tag{9}$$

where $n$ is the total number of nodes, $d(j,i)$ represents the shortest distance between nodes $j$ and $i$, while the term $n-1$ is used to normalize the closeness value and make it comparable across networks of different sizes. In Eq. (9), $NS_I, CL$ indicates Network Structure measure of Indirect linkages based on CLoseness centrality.

*Distinctiveness centrality* builds upon degree centrality but considers the characteristics of the nodes connected to node $i$. Unlike degree centrality, which assigns equal importance to all connections, distinctiveness centrality emphasizes distinctive connections between nodes by penalizing the links to nodes that have a high degree (Fronzetti Colladon and Naldi, 2020). From this perspective, it would be more beneficial for a technology sector to be connected to sectors with fewer connections, than to those with numerous links in the network, as it would allow to benefit exclusive technology transfers. For node $i$, the Distinctiveness index is computed as:

$$L_i^{NS_I,DI} = D(i) = \sum_{j=1, j\neq i}^{n} \log_{10} \frac{n-1}{g_j} I(w_{ij} > 0) \tag{10}$$

where $n$ is the total number of nodes, $g_j$ is the degree of node $j$ and $I(w_{ij} > 0)$ is a function that assumes the value of one if there is an arc connecting nodes $i$ and $j$ with a weight greater than zero. The Distinctiveness index can be normalized by scaling it on its theoretical upper bound, that is $(n-1)\log_{10}(n-1)$. In Eq. (10), $NS_I, DI$ stands for Network Structure measure of Indirect linkages based on DIstinctiveness centrality.[8]

[8] The Distinctiveness formula can also be generalized for directed networks to calculate in- and out-distinctiveness (Fronzetti Colladon and Naldi, 2020).

We condense the information conveyed by the above-mentioned centrality metrics by extracting a common *latent factor* from all these indicators through Principal Component Analysis (PCA), $L^{NS,LF}$ (Lanjouw and Schankerman, 2004). By exploiting information on multiple characteristics of the network, the composite indicator is able to capture common variation across the different centrality measures and leave out idiosyncratic measurement errors, better measuring the intensity of structural linkages within the innovation network. The regression results yielded using the composite factor should be compared with those obtained with the Katz metric built for the overall set of inter-sectoral linkages (namely, direct linkages plus indirect linkages). Computationally, for each year of our timeframe, we build a latent factor based on the first principal component extracted. The factor exploiting information on text similarity explains between 90 and 98% of the variability of centrality metrics. The information content of this index has significantly increased over time. The latent factor built on citation networks is able to explain even a higher portion of variation and possesses an information content which is quite stable over time. In text similarity networks, distinctiveness contributes to the latent common factor more than any other centrality measure, accounting for between 40% and 50% of its variation over time. On the other hand, degree centrality emerges as the primary contributor to the latent factor derived from citation networks, making up approximately 50% of the total variance.

### 4.3. Variable description

We conduct our analysis considering a panel sample of 128 technology sectors identified at the 3-digit level of the CPC classification. The work covers the period from 1976 to 2021. Using various information included in the USPTO dataset, we are able to build three different groups of variables: *innovation outcome*, *technological interdependence* (neighbor innovativeness and network structure) and *control variables*.

As a baseline measure of innovation outcome ($\Delta N$), we utilize the simple count of patent applications. However, to account for heterogeneity in the quality of innovations, we weigh patent counts with the number of citations received (forward citations). Primarily, we adopt an *univocal* assignment approach and attribute each patent to only one technology sector, identified as the first CPC class listed in the patent document. This means that each patent is associated with only one primary class, even though it may be considered as a realization of different technology areas in light of the full list of CPC codes reported in the document. In our robustness checks, we also explore a *multiple* assignment approach. Each patent is hence considered as having multiple realizations and is evenly assigned to all its CPC classes, including primary and secondary classes. When conducting these robustness regressions, we maintain the structure of linkages derived using the univocal approach and exclude any connections between a patent's primary and secondary classes. This helps avoid spurious interdependence among technology sectors.

Technological interdependence is sourced by the pool of knowledge directly related to the degree of innovativeness of neighborings, and to the structure of connections among sectors in the innovation network. *Neighbor innovativeness* is measured with several proximity-weighted averages of innovations developed by other sectors. As discussed above, innovation output is measured both in terms of simple patent counts and forward cites-adjusted number of patent applications. To gauge inter-sectoral knowledge transmission, we evaluate innovation from other sectors with weights extracted from our novel matrix of text (cosine) similarity, as well as from a more traditional matrix reflecting bilateral citations flows. The latter expresses the number of citations made by sector $i$ to sector $j$ as a portion of the total number of citations made by the citing sector. On this basis, our proxies for neighbor innovativeness include the following variables: bilateral Cites-weighted Counts (CWC), bilateral Cites-Weighted Forward cites (CWF), Text similarity-Weighted Forward cites (TWF). The influence of *Network*

**Table 1**
List of the variables.

| Label | Description | Formula |
|---|---|---|
| **Innovation outcome** | | |
| $\Delta N_t$ | Number of raw or quality-adjusted patent counts | |
| **Internal innovation factors** | | |
| $N$ | Cumulative number of raw or quality-adjusted patent counts | $N_{it} = \Delta N_{it} + (1 - 0.15) \times N_{it-1}$ |
| $R$ | Number of inventors per firm's patents | |
| $Z$ | Number of applicants per sector | |
| **Technological interdependence** (time subscript omitted) | | |
| | **Neighbor Innovativeness** ($NI$) | |
| $L^{NI}$ | Bilateral Cites-weighted Counts (CWC) | $L_i^{NI} = \sum_{j=1}^{n} c_{ij} \Delta N_j \quad c_{ij} = \sum_j^n c_j \quad c_{ii} = 0$ |
| | Bilateral Cites-Weighted Forward cites (CWF) | |
| | Text similarity-Weighted Forward cites (TWF) | $L_i^{NI} = \sum_{j=1}^{n} w_{ij} \Delta N_j \quad w_{ii} = 0 \quad w_{ij} = V_i V_j$ |
| | **Network Structure** ($NS$) | |
| $L^{NS_D,KZ}$ | Katz (direct) | $L_i^{NS_D,KZ} = \iota \sum_{j=1}^{n} w_{ji}$ |
| $L^{NS_I,KZ}$ | Katz (indirect) | $L_i^{NS_I,KZ} = L_i^{NS,KZ} - \eta(\iota \sum_{j=1}^{n} w_{ji})$ |
| $L^{NS,KZ}$ | Katz (total) | $L_i^{NS,KZ} = \sum_{l=1}^{\infty} \sum_{j=1}^{n} \eta^l w_{ji}^l$ |
| $L^{NS_D,DG}$ | Degree (direct) | $L_i^{NS_D,DG} = \sum_{j=1,j\neq i}^{n} I(w_{ij} > 0)$ |
| $L^{NS_I,BE}$ | Betweenness (indirect) | $L_i^{NS_I,BE} = \sum_{j<k} \frac{d_{jk}(i)}{d_{jk}}$ |
| $L^{NS_I,CL}$ | Closeness (indirect) | $L_{it}^{NS_I,CL} = \frac{n-1}{\sum_j d(j,i)}$ |
| $L^{NS_I,DI}$ | Distinctiveness (indirect) | $L_{it}^{NS_I,DI} = \sum_{j=1,j\neq i}^{n} \log_{10} \frac{n-1}{g_j} I(w_{ij} > 0)$ |
| $L^{NS,LF}$ | Latent Factor (total) | |

*structure*, which depends on the position of the sector within the technology space, is evaluated through the Katz metrics (for overall, direct and indirect linkages), the other four metrics of network centrality, i.e., Degree, Betweenness, Closeness, and Distinctiveness, as well as the Latent factor derived from the last group of indicators. All measures of technological interdependence are built by exploiting information both on patent text similarity and bilateral citation flows.

Finally, we consider several control variables, proven to affect the outcome of innovation activities in the earlier literature. One of such variables is the cumulative value of internal knowledge, defined as the sector's stock of patents, $N$. The patent stock is built from the annual flow of patent applications with the perpetual inventory method using a depreciation rate of 15%. The same procedure is used when we use quality-adjusted (citation-weighted) measures of patent counts. As a measure of innovation effort, we look at the average number of inventors involved in patenting, computed at the level of the individual firm (applicant) active in each sector. This can be considered as a proxy for the amount of human resources allocated to innovation processes. Lastly, we measure the effect of product diversification with the number of applicants active in each sector. The full list of the variables used in the regression analysis and the methods adopted in their construction are illustrated in Table 1.

As discussed above, we estimate the regression mainly as a log-linear model. To this aim, we handle zeros of the dependent variable using the following transformation, $\ln(1 + \Delta N)$, and then, in robustness checks, assess how this assumption influences the regression results.

### 4.4. Descriptive analysis

The main summary statistics for the variables used in our work are displayed in Table 2, while the matrix of their correlation coefficients can be found in Table A.1 of Appendix. For the sake of brevity, we report means and standard deviations only for the measures of technological interdependence based on text similarity.

On average, over 1100 patents are applied annually by each of the 128 technology sectors, based on univocal patent assignment to primary technology classes. As known, the distribution of patent realizations is very skewed, and the standard deviation of this variable is much larger than its mean (3.5 thousand). The number of forward citations per sector is 796, implying that each application has less than one citation, on average. There is significant variation in the distribution of citations, with some patents receiving a high number of citations and most of them only a few. The standard deviation of forward citations is 2215 per year.

The measures of neighbor innovativeness reveal that the pool of external knowledge potentially available to each sector for implementing its innovation activities is much higher when using a weighting scheme based on patent citation flows (CWC and CWF). However, it can be seen that variation in neighbor innovativeness is much smaller compared to the mean when we use cosine similarity to weigh quality-adjusted patents (TWF). This discrepancy can be attributed to the higher skewness of the citation distribution as opposed to text similarity.

The intensity of structural linkages featuring the innovation network denotes a less uneven distribution compared to neighbor innovativeness. Notably, the standard deviation of all the network centrality measures is smaller than their mean, except for Betweenness, which is highly skewed, and the latent factor which has a standard normal distribution by construction. The mean of the Katz index is 0.02 for direct connections and 23.21 for indirect ones. When considering the other indexes reflecting structural linkages, that are built using information on text similarity, one has to bear in mind that all these measures are normalized (i.e., ranging between 0 and 1), and that links in the lowest quartile of the similarity distribution have been removed. Overall, text similarity networks are quite dense (Mean, M, 0.749, Standard Deviation, SD, 0.048), with a high clustering coefficient (M 0.893, SD 0.021) and a low (unweighted) average shortest path length (M 1.217, SD 0.046). The average normalized degree is 0.746 (SD 0.241), while

**Table 2**
Descriptive Statistics.

| Variable | Mean | SD |
|---|---|---|
| **Innovation outcome** | | |
| Patent counts (univocal) | 1,103.6 | 3,547.4 |
| Forward cites (univocal) | 796.23 | 2,215.8 |
| **Internal innovation factors** | | |
| Cumulative knowledge (patent stock) | 5,929.8 | 17,532.3 |
| Innovation effort (# of inventors per firm) | 2.288 | 0.535 |
| Product proliferation/diversification (# of classes per firm) | 203.9 | 490.4 |
| **Technological interdependence** | | |
| **Neighbor Innovativeness** ($NI$) | | |
| Bilateral Cities-Weighted Counts (CWC) (in 1,000) | 588.7 | 4,838.3 |
| Bilateral Cites-Weighted Forward cites (CWF) (in 1,000) | 240.3 | 1,376.2 |
| Text similarity-weighted forward cites (TWF) (in 1,000) | 22.67 | 14.62 |
| **Network Structure** ($NS$) | | |
| Katz (direct) | 0.023 | 0.007 |
| Katz (indirect) | 23.21 | 7.002 |
| Katz (total) | 22.98 | 7.072 |
| Degree | 0.746 | 0.241 |
| Betweeness | 0.001 | 0.004 |
| Closeness | 0.171 | 0.047 |
| Distinctiveness | 0.032 | 0.014 |
| Latent factor | 0.001 | 0.970 |

**Notes**: Statistics are computed over sectors and years. All measures of technological interdependence use information on text similarity.

the average total similarity score of each sector (average weighted degree) is 20.809 (SD 8.616). This information on the network structure is not reported in Table 2 for brevity.

In Figs. 1 and 2, we show the heatmaps of the pairwise correlation across sectors in terms of citation flows and text similarity, obtained considering the entire time span between 1976 and 2021. The full list of 128 technology classes (3-digit level) is reported on the bottom horizontal and the right-hand vertical axes. The corresponding 2-digit classes (30 sectors) are listed on the top horizontal and the left-hand vertical axes to facilitate the comparison with earlier studies using data at a lower level of disaggregation. Note that we disregard self-citations and self-similarity by setting the value of the cells on the principal diagonal of the matrices to zero. Similarly to Acemoglu et al. (2016b), we normalize the cells of the two matrices on the total sum of each row. In this way, we have row percentages that are fully comparable to the two-digit citation representation provided by Acemoglu et al. (2016b). A few key points are in order. First (and reassuringly), looking at the citation heatmap, there emerges a strong correspondence between our matrix and that reported in the above-cited paper, as denser regions emerge in similar areas of the technology space. Second, comparing our heatmaps, it emerges that the areas with a stronger tone fall in the same technology classes, namely, along the principal diagonal and on the bottom-right cells of the matrix. However, as known, citations concentrate in a few key areas (cells). Conversely, text similarity values are more sparse and homogeneous. This indicates that although our measures of citation and text similarity are likely to capture the same key technological trends, the latter measures are also able to collect information on a broader set of technical characteristics that are more pervasive and, possibly, less technically complex.

## 5. Regression results

### *5.1. Influence of neighbor innovativeness*

We start the analysis by considering the impact of neighbor innovativeness on sectoral innovation capacity (Table 3). As discussed above, we relate to the literature on knowledge spillovers where these transfers are measured in terms of the innovation output of neighboring sectors, weighted by a proximity measure between the sourcing and recipient entities. Our primary interest is to evaluate whether the effect of this source of technological interdependence changes with the nature of the patent variable used to quantify innovation output (namely, patent counts vs forward cites-weighted patents) and of the information used to track intersectoral linkages (namely, bilateral citation flow vs patent text). In this section, we will illustrate how far our estimates fall from the major results of this literature.

In our starting regression (column (1)), we measure innovation output in terms of patent counts per univocal technology sector (i.e., each patent is assigned to its primary class), and neighbor innovativeness in terms of the citation-weighted mean of innovations patented by the other sectors of the economy (CWC). This regression shows that sector interdependence, induced by neighbor innovativeness, is positively and significantly related to the innovation output of the other technology areas.[9] Quantitatively, a one-percent increase in the innovation output of linked (sourcing) sectors is associated with a 0.122 percent increase in the patenting performance of recipient sectors. Comparable evidence for the US can be found, among others, in Jaffe (1986), while more recent estimates provided by Bloom et al. (2013) lie at an upper bound (around 0.4). As known, the count of patent applications as a measure of innovation output is not informative about the quality of new technologies. Hence, in column (2), we do weigh each realized innovation (patent count) with the number of citations received (CWF), finding a larger effect for our proxy for neighbor innovativeness (0.151).

In column (3), we run the previous regression adding as a further explanatory variable the pool of external knowledge made available by neighboring innovators measured using the proximity matrix based on patent text similarity (TWC). This regression highlights that both proxies for neighbor innovativeness are statistically significant and quantitatively important, suggesting that they may capture two distinct dimensions of knowledge transfers. While citations may trace intersectoral links around (parts of) key technologies, patent texts could capture connections across a wider range of technical features. The second dimension of knowledge transfers, measured exploiting information extracted from patent documents, has been broadly neglected in the literature but, quantitatively, it looks as important as that identified by patent citations, as discussed in Feng (2020).

[9] In this (and following) regression(s), we observe that the adjustment parameter is always negative and statistically significant, indicating the existence of a stable (equilibrium) relationship between dependent and explanatory variables in the long run.

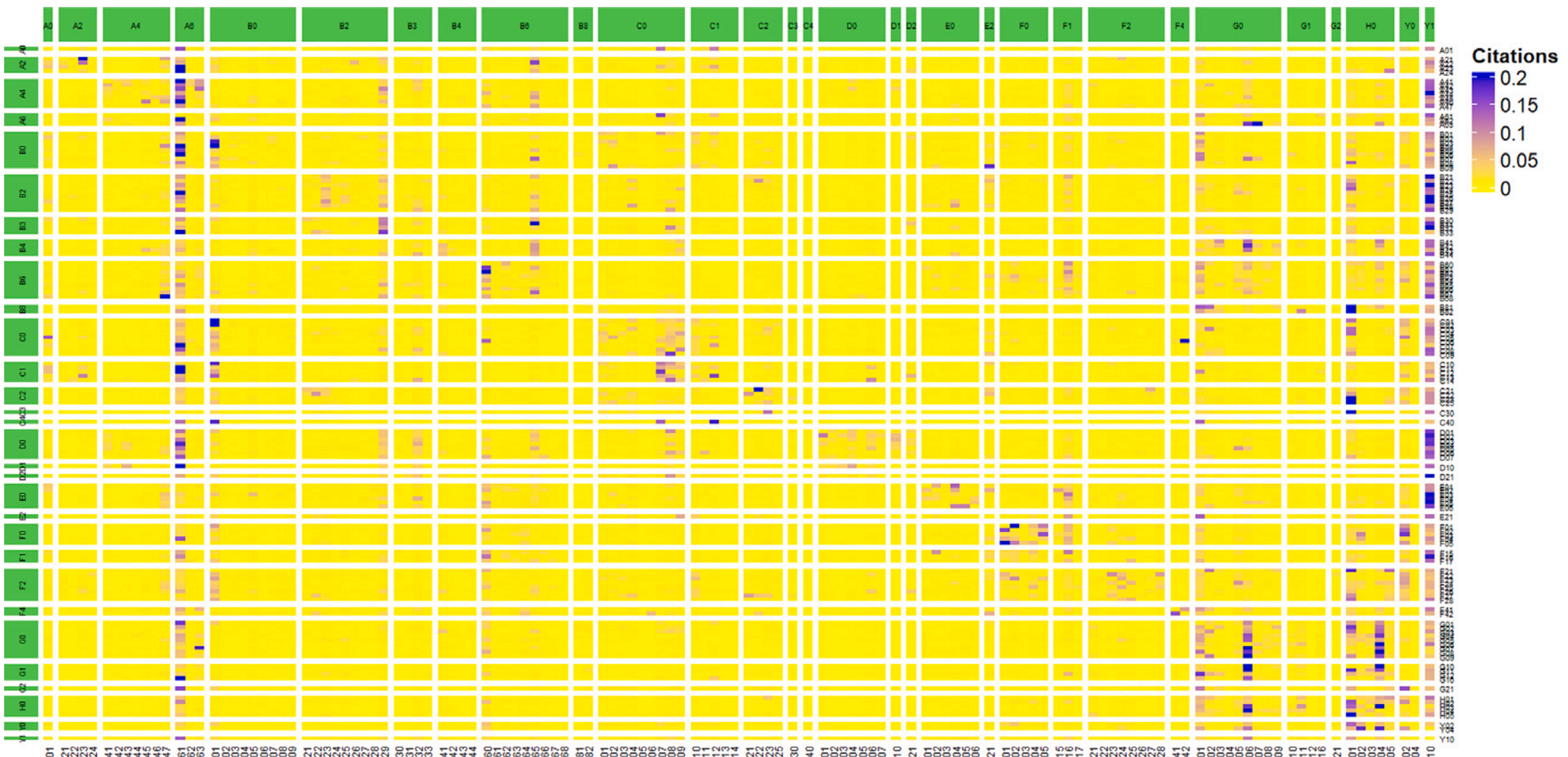

**Fig. 1.** Heatmap of citations.
**Notes**: CPC 1 - and 2 digit categories. A-Human Necessities A0-Agriculture. A2-Foodstuffs; Tobacco. A4-Personal or Domestic Articles. A6-Health; Life-Saving; Amusement. A9-Miscellaneous, of Human Necessities. B-Performing Operations; Transporting. B0-Separating; Mixing. B2-Shaping. B3-Shaping. B4-Printing. B5-Transporting. B6-Microstructural Technology; Nanotechnology. B9-Miscellaneous, Of Performing Operations; Transporting. C-Chemistry; Metallurgy. C0Chemistry. C2-Metallurgy. C3-Metallurgy. C4-Combinatorial Technology. C9-Miscellaneous, of Chemistry; Metallurgy. D-Textiles; Paper. D0-Textiles or Flexible Materials not Otherwise Provided for. D2-Paper. D9-Miscellaneous, of Textiles; Paper. E-Fixed Constructions. E0-Building. E2-Earth or Rock Drilling; Mining. E9-Miscellaneous, Of Fixed Constructions. F-Mechanical Engineering; Lighting; Heating; Weapons; Blasting. F0-Engines or Pumps. F1-Engineering in General. F2-Lighting; Heating. F4-Weapons; Blasting. F9-Miscellaneous, of Mechanical Engineering; etc. G-Physics. G0-Measuring; Optics; Horology; Controlling; Computing; Signaling. G1-Acoustics; Information Storage; Instruments; ICT Adapted to Applications. G2-Nuclear Physics; Nuclear Engineering. G9-Miscellaneous, of Physics.
Each value in the cells is row normalized.

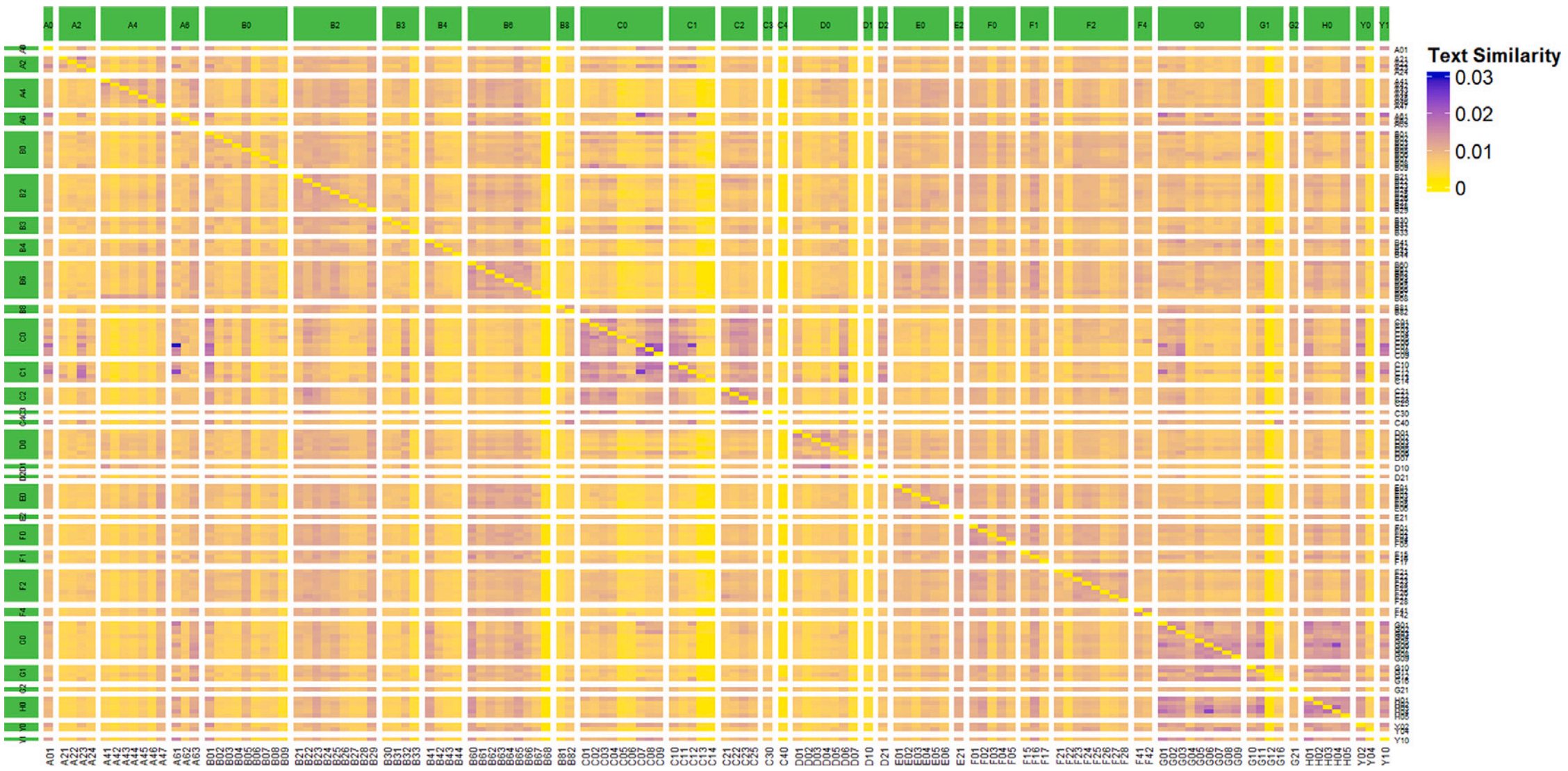

**Fig. 2.** Heatmap of text similarity.
**Notes**: CPC 1- and 2 digit categories. A-Human Necessities A0-Agriculture. A2-Foodstuffs; Tobacco. A4-Personal or Domestic Articles. A6-Health; Life-Saving; Amusement. A9-Miscellaneous, of Human Necessities. B-Performing Operations; Transporting. B0-Separating; Mixing. B2-Shaping. B3-Shaping. B4-Printing. B5-Transporting. B6-Microstructural Technology; Nanotechnology. B9-Miscellaneous, Of Performing Operations; Transporting. C-Chemistry; Metallurgy. C0Chemistry. C2-Metallurgy. C3-Metallurgy. C4-Combinatorial Technology. C9-Miscellaneous, of Chemistry; Metallurgy. D-Textiles; Paper. D0-Textiles or Flexible Materials not Otherwise Provided for. D2-Paper. D9-Miscellaneous, of Textiles; Paper. E-Fixed Constructions. E0-Building. E2-Earth or Rock Drilling; Mining. E9-Miscellaneous, Of Fixed Constructions. F-Mechanical Engineering; Lighting; Heating; Weapons; Blasting. F0-Engines or Pumps. F1-Engineering in General. F2-Lighting; Heating. F4-Weapons; Blasting. F9-Miscellaneous, of Mechanical Engineering; etc. G-Physics. G0-Measuring; Optics; Horology; Controlling; Computing; Signaling. G1-Acoustics; Information Storage; Instruments; ICT Adapted to Applications. G2-Nuclear Physics; Nuclear Engineering. G9-Miscellaneous, of Physics.
Each value in the cells is row normalized.

It should be noted that, thus far, we have adopted a parsimonious algorithm that disregards word frequency in constructing the measure of text similarity (see Section 4 for details). This, however, could artificially increase the similarity between the abstracts. For this reason, we replicate our benchmark regression in column (3) using a text similarity matrix derived by considering all word occurrences, finding similar results. This and all the other results which we cite hereinafter, but that are unreported for the sake of brevity, are available upon request.

**Table 3**
Long-run estimates for the effect of Neighbor Innovativeness.

| | (1) | (2) | (3) | (4) | (5) | (6) |
|---|---|---|---|---|---|---|
| **Neighbor Innovativeness** | | | | | | |
| Citation flows | 0.122*** | 0.151*** | 0.107*** | 0.098*** | 0.089*** | 0.009*** |
| | (0.002) | (0.001) | (0.002) | (0.010) | (0.004) | (0.001) |
| Text similarity | | | 0.541*** | 0.519*** | 0.661*** | 0.117*** |
| | | | (0.0113) | (0.010) | (0.021) | (0.023) |
| **Controls** | | | | | | |
| Cumulative internal knowledge | | | | | | 0.882*** |
| | | | | | | (0.051) |
| Innovation effort | | | | | | 0.031 |
| | | | | | | (0.052) |
| Technology diversification | | | | | | 0.169*** |
| | | | | | | (0.014) |
| **Adjustment parameter** | −0.081*** | −0.069*** | −0.101*** | −0.101*** | −0.101*** | −0.359*** |
| | (0.009) | (0.008) | (0.010) | (0.010) | (0.010) | (0.055) |
| Patent variable | Patent counts | Forward cites | Forward cites | Forward cites | Forward cites | Forward cites |
| Matrix weights | Cites | Cites | Text | Text | Text | Text |
| | (CWC) | (CWF) | (TWF) | (TWF) | (TWF) | (TWF) |
| Patent assignment | Univocal | Univocal | Univocal | Multiple | Univocal | Univocal |
| Sector aggregation (CPC) | 3 digit | 3 digit | 3 digit | 3 digit | 2 digit | 3 digit |
| Observations | 5,632 | 5,632 | 5,632 | 5,632 | 1,320 | 5,390 |
| R-squared | 0.145 | 0.146 | 0.141 | 0.141 | 0.090 | 0.041 |
| Number of sectors | 128 | 128 | 128 | 128 | 30 | 128 |

**Notes**: Long-run estimates (elasticities) derived from an ARDL(2,1). All variables are expressed in logs. Heteroskedasticity-Autocorrelation Consistent (HAC) standard errors are in parentheses. All regressions use sector-specific fixed effects and account for the effect of common time shocks (time dummies) using variables expressed in deviation from their yearly means. Innovation is measured by the raw number of patent counts (column (1)) and by the forward cites-adjusted number of patent counts (columns (2)–(6)). The matrix of technological proximity is based on bilateral citation flows in columns (1)–(2), and on pairwise cosine similarity of patent texts in columns (3)–(6). Each patent is univocally assigned to one technology sector (class) in columns (1)–(3) and (5)–(6), and to multiple sectors based on the full list of technology classes listed in the patent document in column (4). Technology classes (sectors) are based on the Cooperative Patent Classification (CPC). Columns (1)–(4) and (6) use data at the 3-digit level of technology classes (128 sectors); column (5) uses data at the 2-digit level (30 sectors). *** , ** , * denotes statistical significance at the 1, 5 and 10% level, respectively.

One issue in the estimates, that we have discussed above, is that each patent is associated with only one (primary) class when it could be seen as the realization of an innovation in different technology areas (sectors), as resulting from the full list of technology classes reported in the patent document. When considering patents assigned univocally to one technology class, we may downstate the technological capabilities of the firm and, in turn, overstate the impact of neighbor innovativeness, as we do not discern the firm capacity to develop technologies in contiguous technology areas. This implies that, in the benchmark regression, our proxy for neighbor innovativeness might capture the effect of horizontal relatedness, rather than that of genuine knowledge transfers. However, one can broadly infer the technological capabilities of the companies by examining all technology classes outlined in the patent documents. For this reason in column (4), we run our regression with a multiple class assignment for each patent, but preserve the structure of technological linkages used previously (i.e., the same citation-based and text-based distance matrices), so to avoid spurious interdependency between sectors. In this regression, the parameter size of both explanatory variables falls only marginally with respect to the benchmark regression.[10]

Another potential concern regarding our estimates is that the effect of neighbors' innovation capacity may be oversized as patent classifications are imperfect and fuzzy demarcations of the actual structure of the knowledge economy. In order to validate the accuracy of our main results, we conduct a regression analysis using data at a less detailed level of disaggregation. Namely, we consider data for 30 sectors at the two-digit level of the CPC categorization (column (5)). Although it is difficult to predict the direction of the bias associated with the measurement errors caused by imperfect patent classification, which would affect both dependent and explanatory variables, the results in column (5) unequivocally confirm the effect of neighbor innovativeness.[11]

Finally, in the last regression of Table 3, we assess our estimates to omitted variables' bias and control for the effect of internal sources of innovative success. Specifically, we include into the regression the cumulative value of innovations developed in the past by the sector (the patent stock), the average amount of innovation resources currently used by the firms (number of inventors per patent), and the degree of firm product diversification/proliferation (number of applicants active in each sector). As column (6) shows, there is a systematic fall in the influence of the neighbors' innovation when including our set of control variables. As discussed above, it is likely to reflect the strong persistence over time in the effects of knowledge spillovers, technology transfers, etc. There is a long-lasting comovement in innovation activities across sectors, implying a strong correlation between the cumulative value of a sector's patents and our measure of neighboring innovativeness. Intertemporal knowledge returns (or dynamic returns) are a typical driver of innovation outcomes (Caballero and Jaffe, 1993): firms with a larger technological knowledge, developed over time through successful innovation, have an advantage in generating new knowledge compared to innovators with a smaller past engagement in R&D. The parameter size of the patent stock (0.882) signals that inter-temporal (within-industry) spillovers are positive but slightly decreasing over time. This may reflect the increasing difficulty of doing R&D caused by diminishing technological opportunities or the fall in the cost-effectiveness of R&D (Bloom et al., 2020). This finding departs from major results of the earlier literature based on cross-country data (from Madsen,

[10] Note that if we assign each application fractionally to all classes reported in the patent document, the fall in the parameter size of neighbor innovativeness is greater than in column (4). This finding should be taken with caution due to the impossibility of assigning patent documents proportionally to various technology classes (sectors). This is likely to generate a classical measurement error that downward biases the parameter of our proxies for neighbor innovativeness.

[11] See Lafond and Kim (2019) for a pioneering application of endogenous clustering to the USPTO data.

**Table 4**
Long-run estimates for the effect of Network Structure: Katz centrality metrics.

| | (1) | (2) | (3) | (4) | (5) | (6) | (7) | (8) |
|---|---|---|---|---|---|---|---|---|
| **Neighbor Innovativeness** | | | | | | | | |
| Text similarity | 0.117*** | | | 0.321*** | 0.112*** | 0.283*** | 0.120*** | 0.233*** |
| | (0.023) | | | (0.061) | (0.022) | (0.051) | (0.022) | (0.056) |
| Citation flows | 0.009*** | | | 0.008*** | 0.007*** | 0.008*** | 0.008*** | 0.007*** |
| | (0.001) | | | (0.001) | (0.001) | (0.001) | (0.001) | (0.001) |
| **Network Structure** | | | | | | | | |
| Katz (total) | | | | | | | | |
| Text similarity | | 0.074*** | | −0.197*** | | | | |
| | | (0.020) | | (0.053) | | | | |
| Citation flows | | | 0.008*** | | 0.006*** | | | |
| | | | (0.001) | | (0.001) | | | |
| Katz (direct) | | | | | | | | |
| Text similarity | | | | | | −0.929*** | | −0.773*** |
| | | | | | | (0.142) | | (0.161) |
| Citation flows | | | | | | | 0.014*** | 0.012*** |
| | | | | | | | (0.005) | (0.005) |
| Katz (indirect) | | | | | | | | |
| Text similarity | | | | | | 0.799*** | | 0.713*** |
| | | | | | | (0.123) | | (0.129) |
| Citation flows | | | | | | | 0.003*** | 0.003*** |
| | | | | | | | (0.001) | (0.001) |
| **Controls** | Yes | Yes | Yes | Yes | Yes | Yes | Yes | Yes |
| **Adjustment par.** | −0.359*** | −0.355*** | −0.376*** | −0.361*** | −0.362*** | −0.361*** | −0.363*** | −0.365*** |
| | (0.055) | (0.056) | (0.059) | (0.055) | (0.055) | (0.058) | (0.055) | (0.058) |
| Obs. | 5,390 | 5,390 | 5,390 | 5,390 | 5,390 | 5,390 | 5,390 | 5,390 |
| R-squared | 0.041 | 0.042 | 0.043 | 0.041 | 0.041 | 0.041 | 0.041 | 0.041 |

**Notes**: Long-run estimates (elasticities) derived from an ARDL(2,1). All variables are expressed in logs. Heteroskedasticity-Autocorrelation Consistent (HAC) standard errors are in parentheses. All regressions account for the effect of common time shocks (time dummies) using variables expressed in deviation from their yearly means. Controls included: Cumulative internal knowledge; Innovation effort; Technology diversification. Innovation output is measured by the forward cites-adjusted number of patent counts. The matrix of technological proximity is based on bilateral citation flows and on pairwise cosine similarity of patent texts. Each patent is univocally assigned to one technology sector. *** , ** , * denotes statistical significance at the 1, 5 and 10% level, respectively.

2008 onwards), which points to highly persistent returns of past knowledge on the creation of innovations (constant intertemporal spillovers). However, our evidence of decreasing returns to scale of R&D aligns to previous studies conducted on the US and European industries (Venturini, 2012; Mason et al., 2020). The impact of the current R&D effort, here approximated by the number of inventors involved in innovative processes, seems to overlap with that of the knowledge (patent) stock, as the former explanatory variable, albeit positively signed, is never significant. While expanding product varieties should depress the net returns to innovation effort according to the Schumpeterian growth theory (Ha and Howitt, 2007), we find a positive effect for our proxy for product proliferation/diversification, namely the number of applicants active in each sector. This would signal that companies active in the same sector are likely to exploit technological complementarities or economics of scope in their innovation processes.

In the Appendix, we conduct a set of econometric checks on our baseline regression (Table A.2). Specifically, we alternatively use: (i) a richer dynamic adjument to neutralize the effect of reverse casuality; (ii) a linear dynamic regression robust to misspecified dynamics and error serial correlation (the Cross-Sectional augmented Distributed-Lag, CS-DL, in place of the ARDL regression); (iii) robust controls to common un-observable factors (Common Correlated Effects, CCE, in place of time dummies); (iv) counterfactual (random) distance weights; (v) count data regression (negative binomial); and finally (vi) inverse hyperbolic sine transformation to account for missing values. In all these cases, our main results of Table 3 are largely confirmed.

### *5.2. Influence of the network structure*

Next, we account for the effect of structural linkages by utilizing a comprehensive range of network centrality measures that have been previously introduced. We present the results of this analysis in two parts. Table 4 reports estimates obtained adopting the Katz metric, which is also decomposed to assess the influence on innovation performance exerted by the direct and indirect connections existing across technology sectors. Table 5 illustrates the results obtained using the index of degree centrality for measuring direct linkages, and separately the indicators of betweenness, closeness, and distinctiveness for gauging indirect linkages. The latter table also presents the results obtained using the latent factor, built to capture the full spectrum of effects produced by structural linkages, as measured by the second group of network centrality indicators. Again, in both regression tables we include measures of technological interdependence constructed using either bilateral citations or text similarity.

In columns (2) and (3) of Table 4, we include the Katz centrality index measuring the overall network of structural linkages and observe that it is positively and significantly related to sectoral innovation output. In line with our earlier estimates, the coefficient size of the variable based on text similarity is much larger than the one obtained using citation flows. However, the coefficient of the Katz index based on text similarity turns negative when we include this variable in the same specification with our measure of neighbor innovativeness (column (4)). This is as the latter regressor is built using (bilateral) direct linkages to weigh the innovation output of sourcing sectors. Consistently,

**Table 5**
Long-run estimates for the effect of Network Structure: Network centrality metrics.

| | (1) | (2) | (3) | (4) | (5) | (6) |
|---|---|---|---|---|---|---|
| **Neighbor Innovativeness** | | | | | | |
| Text similarity | 0.117*** | 0.040* | 0.205*** | 0.074*** | 0.087*** | 0.067*** |
| | (0.023) | (0.021) | (0.024) | (0.021) | (0.022) | (0.020) |
| Citation flows | 0.009*** | 0.008*** | 0.009*** | 0.008*** | 0.008*** | 0.008*** |
| | (0.001) | (0.001) | (0.001) | (0.001) | (0.001) | (0.001) |
| **Network Structure** | | | | | | |
| Degree | | | | | | |
| Text similarity | | 0.001*** | | | | |
| | | (0.001) | | | | |
| Citation flows | | 0.012*** | | | | |
| | | (0.001) | | | | |
| Betweeness | | | | | | |
| Text similarity | | | 0.004 | | | |
| | | | (0.003) | | | |
| Citation flows | | | 0.011*** | | | |
| | | | (0.001) | | | |
| Closeness | | | | | | |
| Text similarity | | | | 0.009** | | |
| | | | | (0.004) | | |
| Citation flows | | | | 0.006*** | | |
| | | | | (0.001) | | |
| Distinctiveness | | | | | | |
| Text similarity | | | | | 0.015*** | |
| | | | | | (0.004) | |
| Citation flows | | | | | 0.006** | |
| | | | | | (0.003) | |
| Latent factor | | | | | | |
| Text similarity | | | | | | 0.041*** |
| | | | | | | (0.009) |
| Citation flows | | | | | | 0.035*** |
| | | | | | | (0.006) |
| **Controls** | Yes | Yes | Yes | Yes | Yes | Yes |
| **Adjustment par.** | −0.359*** | −0.319*** | −0.314*** | −0.376*** | −0.369*** | −0.631*** |
| | (0.055) | (0.052) | (0.050) | (0.055) | (0.050) | (0.035) |
| Obs. | 5,390 | 5,390 | 5,390 | 5,390 | 5,390 | 5,390 |
| R-squared | 0.041 | 0.041 | 0.041 | 0.041 | 0.041 | 0.041 |

**Notes**: Long-run estimates (elasticities) derived from an ARDL(2,1). All variables are expressed in logs, except Degree, Betweenness, Closeness, Distinctiveness, and the latent factor which enter the regression multiplied by 100 so that their parameters can be treated as elasticities. Heteroskedasticity-Autocorrelation Consistent (HAC) standard errors are in parentheses. Controls included: Cumulative internal knowledge; Innovation effort; Technology diversification. Innovation output is measured by the forward cites-adjusted number of patent counts. The matrix of technological proximity is based on bilateral citation flows and on pairwise cosine similarity of patent texts. Each patent is univocally assigned to one technology sector. *** , ** , * denotes statistical significance at the 1, 5 and 10% level, respectively.

when we decompose the Katz metric into the effects associated with direct and indirect linkages, the former are found to be negatively related to innovation output, while the latter variable has a positive coefficient (column (6)). This finding clearly points to the overlapping between the text-based measure of neighbor innovativeness and the text-based measure of direct linkages. It is important to emphasize that a different pattern of results is found when the Katz measures are derived from the networks of citation flows (columns (7) and (8)): these variables are always positively and significantly associated with sector innovation and, albeit small in size, present quite stable coefficients across regressions (see Taalbi, 2020 for consistent results).

In Table 5, we explore the influence of the network structure using our second group of centrality indicators. First, we extend the benchmark regression (with controls) with a measure of direct linkages captured by degree centrality (column (2)).[12] The effect of this explanatory variable is positive and significant both when it is based on text similarity and on citation flows. However, the impact of the former version of the variable largely overlaps with the effect of neighbor innovativeness, which uses direct linkages to weigh external innovation. Consequently, the coefficient of the variable capturing knowledge spillovers falls at the limit of the significance region in column (2). At the same time, the impact associated with the direct connections within the network, as measured by degree centrality based on bilateral citations, is statistically significant and economically substantial.

As a second step, we quantify the impact of the overall linkages channeled by the complex structure of intersectoral relations (columns (3)–(6)). We find strong indication that the network structure positively affects innovation performance with all indicators adopted, expect when using the index of closeness centrality calculated on text similarity. It is important to highlight that the long-run elasticity identified for the latent factor (column (6)) largely exceeds the impact estimated for each individual measure of centrality from which it is derived (columns (2)–(5)). This suggests that the latent factor has the ability to capture variation in structural linkages within the network that would otherwise go unnoticed when utilizing individual measures of network centrality one at a time.

Summing up, the results of this section reinforce the idea that the network structure (topology) of sectoral linkages is both complex and relevant, and that social network analysis methods can effectively help leverage its informative content. More importantly, evidence in Tables 4 and 5 suggests that the effect of network linkages on the success of sectoral innovation activities is comparable in size to that of neighbor innovativeness.

12 In Table 5, all variables are expressed in logs, except Degree, Betweenness, Closeness, Distinctiveness, and the latent factor. These variables enter the regression multiplied by 100 so that their parameters can be treated as elasticities and are comparable to the coefficient of the other regressors.

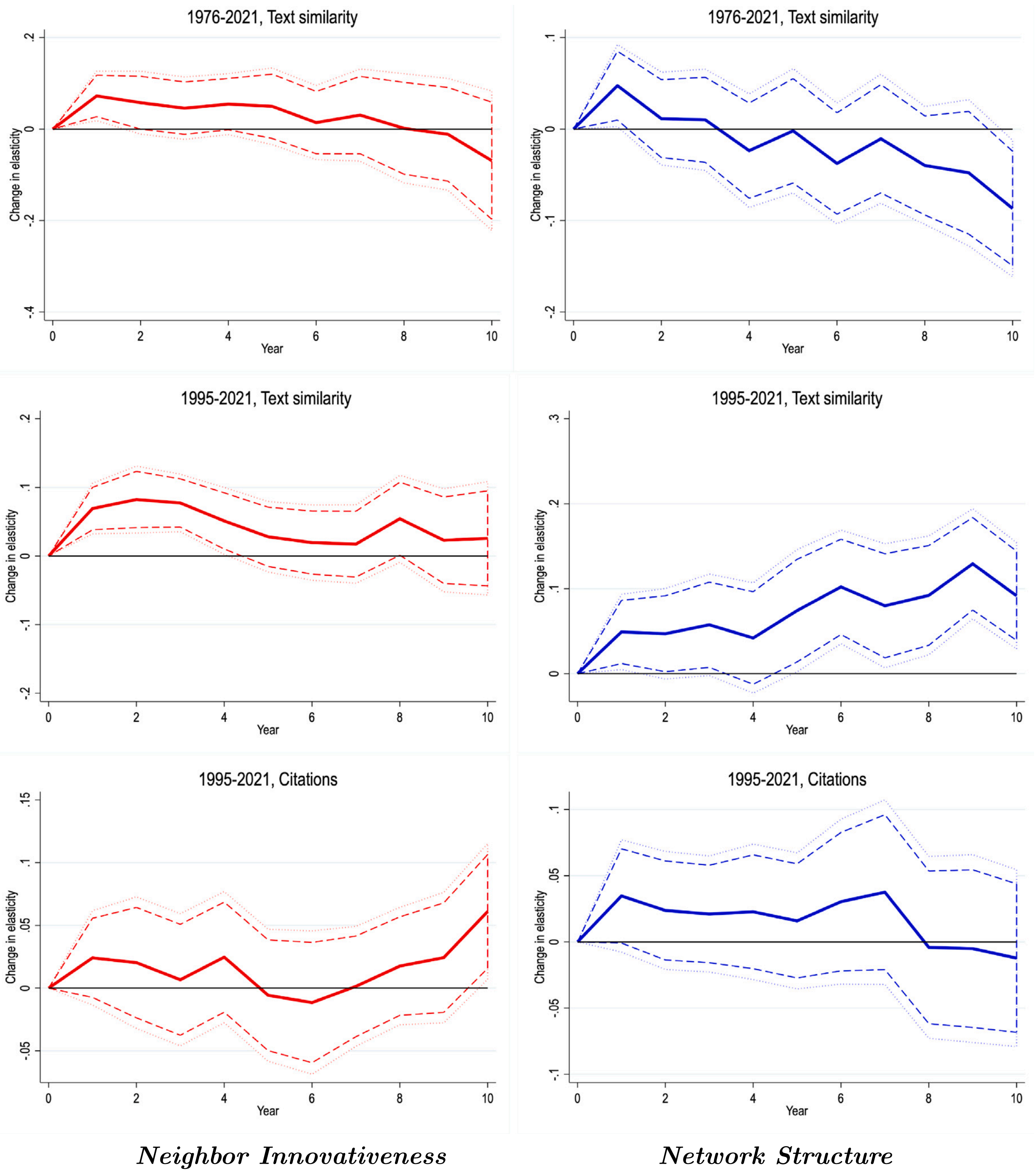

**Fig. 3.** Innovation response to technological shocks: Neighbor innovativeness vs network structure.
**Notes**: The graph reports the forward effect coefficients associated with the local projection estimation in Eq. (11). All estimates use fixed effects, time dummies, and standard errors robust to heteroskedasticity. The graphs report the confidence bands at 90 and 95%. The event is identified with the year of peak increase in the variable under assessment. The graph on the left-hand (right-hand) side considers a shock to neighbor innovativeness (network position) while keeping the other measure of technological interdependence as control. The degree of neighbor innovativeness is measured with the proximity-weighted number of quality-adjusted patents of the other sectors. The structure of the network linkages is summarized by the latent factor extracted from the indicators of network degree, betweenness, closeness and distinctiveness centrality.

### *5.3. Innovation response to technological shocks in linked sectors*

We conclude the analysis by assessing how shocks, that affect technological interdependence, spread through the technology space and affect sector innovation. Specifically, we implement an event analysis and simulate the change in innovation output (response) after an innovation shock in technologically related sectors (impulse). We use our dynamic regression model to perform a local projection analysis as originally devised by Jordà (2005).[13] For each sector, we consider as an event the year with the peak increase in innovation outcome of connected sectors, $L^v$ ($v = NI, NS$), and assess the associated change in $\Delta N$ within the horizon of ten years after the shock. In essence, we run a set of forward-effect regressions in which the event variable is a dummy of value one in the peak year, and zero otherwise. For all regressors, we consider a set of lags, $p$, to filter out the effect of their adjustment dynamics and a set of leads to exclude bias induced by anticipation effects ($p = 2$). The specification used in the event analysis is shaped as follows:

$$\Delta N_{i,t+k} = a_i + \sum_{p=1}^{3} a_1 \Delta N_{i,t-p} + \sum_{p=0}^{2} a_2 E_{i,t-p} + \sum_{p=0}^{2} a_3 X_{i,t-p} + \sum_{h=0}^{k} (a_4 E_{it+h} + a_5 X_{i,t+h}) + \epsilon_{it} \quad (11)$$

where $E$ identifies the event dummy, $X$ are the control variables, and $\epsilon_{it}$ the error terms. $k = 1, \ldots, 10$ is the time horizon of the estimated forward effect. All estimations use fixed effects, time dummies, and standard errors robust to heteroskedasticity. The graphs report the confidence bands at 90 and 95%.

[13] See Akcigit et al. (2022) and Madsen et al. (2024) for recent applications in macro-economic analysis of innovation.

We consider two types of shocks separately, one on the variable measuring the degree of neighbors' innovativeness, and one on the variable reflecting the intensity the structural linkages, namely the latent factor. We treat as primary proxy for technological interdependence the measure constructed on text similarity and use the other variable, based on bilateral citations, as a control. However, as a counterfactual exercise, we look at the peak increase (shock) in the citation-based measure of technological interdependence and employ the other measure based on text similarity as a control. Since we are interested in understanding whether the propagation of technological shocks has changed recently, we distinguish the responses by the time interval of the data. Accordingly, we first consider the entire period between 1976 and 2021 and then look at the interval since 1995.[14]

The results of the event analysis offer particularly valuable insights (see Fig. 3). Considering the overall time span, there is no systematic response of sector innovation to positive shocks affecting both dimensions of technological interdependence (neighbor innovativeness and network structure). However, the pattern of results changes remarkably from 1995 onward, when the effect of an unanticipated change in neighbor innovativeness turns out to be significant and manifests sooner than shocks affecting the network structure. The effect of the former shocks emerges in (and remains significant for) less than five years, while the impact of an unexpected change in network linkages becomes significant after half decade and grows smoothly over time. Our simulation results suggest that the influence of network linkages may even prevail for magnitude over the impact of neighbor innovativeness.

Lastly, in order to ascertain whether patent documents are a valid source of information on technological shocks affecting sector innovation, we perturb the citation-based measures of technological interdependence, considering as an event the year with a peak increase in this group of variables. In such a regression, we do not find any appreciable response in patenting activities of the sectors. This finding confirms that patent text conveys valuable information on the development of new technologies and on the direction of technological change.

[14] As argued above, we account for the effect of time shocks, having a homogeneous effects across sectors, using time dummies. This type of regressions do not support the control for the impact of cross-sector heterogeneous effects through CCEs. However, the risk that unobservable ("third") factors are driving the results of the event analysis is excluded by the fact that only 3% of (4 out 128) sectors has the same event year (i.e., the peak year in the impulse variable).

## 6. Conclusions

This paper has analyzed how technological interdependence affects the sectors' ability to innovate, looking at the effects of neighbor innovativeness and at the structure (topology) of technological linkages within the innovation network. By examining the abstract of 6.5 million patents, granted by the United States Patent and Trademark Office (USPTO) between 1976 and 2021, we have first shown that both dimensions of technological interdependence matter for sectoral innovation. In the long run, the impact of network linkages turns out to be as important as that of neighbor innovativeness. In the relatively short run, positive shocks to neighbor innovativeness produce relatively faster effects, while the impact of the shocks strengthening network linkages lasts longer and is apparently quantitatively larger. We have then demonstrated that patent texts represent a rich source of information on innovations' content. On average, text-based data exhibit a greater level of consistency compared to cite-based data and are suitable to identify both the effect of neighbor innovativeness and network structure.

All these findings have important implications for the evolution of research on innovation and technological change, knowledge spillovers and structural (network) linkages. Our study has highlighted that the natural language processing of patent documents offers a unique opportunity to move beyond traditional methods of innovation classification and analysis, such as the CPC categorization. By employing text mining techniques, researchers can endogenously determine clusters of similar technologies, can study how these evolve over time and uncover factors behind their development. Another area deserving exploration is the analysis of patent text (dis)similarity for the purpose of technological forecasting, i.e., to predict which areas of the technology space are likely to expand in the near future. A newer generation of textual analysis techniques, for example based on transformers or large language models (ChatGPT, etc.), could be used to this purpose in light of their high potential. There has been increasing effort in social sciences, especially among sociologists, in learning human behavior through digital tracks collected via "cognitive artifacts" (e.g., smartphones). Large language models such as those based on transformers are increasingly important for the develop the new lines of research, for instance as those focused on the interplay between humans and ambient intelligence (the Metaverse). This could help understand the latest evolution of the technology space towards new digital fields (Godwin-Jones, 2021 and 2023). These topics are left as subjects for future research.

### CRediT authorship contribution statement

**Andrea Fronzetti Colladon:** Writing – review & editing, Writing – original draft, Visualization, Software, Methodology, Formal analysis, Conceptualization. **Barbara Guardabascio:** Writing – review & editing, Writing – original draft, Methodology, Visualization, Formal analysis, Data curation. **Francesco Venturini:** Writing – review & editing, Writing – original draft, Methodology, Formal analysis, Data curation, Conceptualization.

### Data availability

Data will be made available on request.

### Acknowledgment

The authors gratefully acknowledge the Editor, Prof. Adam Jaffe, and three anonymous referees for helpful comments.

**Table A.1**
Correlation structure.

| Label | Counts | FWC | CWC | CWF | TWF | Katz direct | Katz indirect |
|---|---|---|---|---|---|---|---|
| Counts | 1.000 | | | | | | |
| FWC | 0.8991* | 1.000 | | | | | |
| CWC | 0.7034* | 0.3727* | 1.000 | | | | |
| CWF | 0.8616* | 0.6706* | 0.8106* | 1.000 | | | |
| TWF | 0.4103* | 0.4721* | 0.1356* | 0.3218* | 1.000 | | |
| Katz direct | 0.2566* | 0.2727* | 0.1223* | 0.1605* | 0.6210* | 1.000 | |
| Katz indirect | 0.2765* | 0.2945* | 0.1325* | 0.1751* | 0.6290* | 0.9993* | 1.000 |
| Katz total | 0.2765* | 0.2945* | 0.1325* | 0.1751* | 0.6290* | 0.9993* | 1.000* |
| Degree | 0.1111* | 0.1232* | 0.0484* | 0.0624* | 0.4648* | 0.8899* | 0.8797* |
| Betweeness | 0.0390* | 0.0684* | −0.0210 | −0.0219 | 0.1809* | 0.3615* | 0.3580* |
| Closeness | 0.1939* | 0.2056* | 0.0906* | 0.1157* | 0.5422* | 0.9760* | 0.9715* |
| Distinctiveness | −0.0169 | 0.0370* | −0.0751* | −0.0769* | 0.2650* | 0.7136* | 0.7052* |
| Latent Factor | 0.1203* | 0.1437* | 0.0372* | 0.0522* | 0.4784* | 0.9380* | 0.9296* |

| Label | Katz total | Degree | Betweeness | Closeness | Distinctiveness | Latent Factor |
|---|---|---|---|---|---|---|
| Katz total | 1.000 | | | | | |
| Degree | 0.8797* | 1.000 | | | | |
| Betweeness | 0.3580* | 0.2566* | 1.000 | | | |
| Closeness | 0.9715* | 0.8996* | 0.3498* | 1.000 | | |
| Distinctiveness | 0.7052* | 0.7748* | 0.4403* | 0.7254* | 1.000 | |
| Latent Factor | 0.9296* | 0.9649* | 0.4211* | 0.9524* | 0.8619* | 1.000 |

**FWC**: Forward cites (univocal). **CWC**: Bilateral cities-weighted patent counts. **CWF**: Bilateral weighted Forward cites. **TWF**: Text similarity-weighted forward cites. All centrality measures are based on text similarity. * significant at 5%

## Appendix

### A.1. Econometric checks

In Table A.2, we display the results of some econometric checks conducted on our baseline regression illustrated in Table 3, which is reported here in column (1) as a reference. As a first check, we estimate our specification using a richer dynamic adjustment. It is known that the ARDL specification yields consistent estimates, which are robust to reverse causality when the lag structure is sufficiently rich, but such estimates may be inefficient if too many lags are used in the regression. For this reason, in columns (2) and (3), we extend the lag order of the variables to further two and four lags, finding an effect for neighbor innovativeness (and related knowledge spillovers) consistent with our earlier regressions.

To further address the sensitivity of our results to the modeling of the dynamic adjustment, we run the Cross-Sectional augmented Distributed-Lag (CS-DL) regression (Chudik et al., 2016). With respect to the ARDL model, this procedure has a superior performance when the dynamic of the variables is misspecified and there is error serial correlation, especially with moderately long data like ours. CS-DL estimates in Column (4) reveal a parameter size for neighbor innovativeness only slightly smaller than ARDL estimates, suggesting that our benchmark estimates in column (1) can be considered highly plausible.[15]

In column (5), we run the model on original data (i.e., all series are not expressed as time-de-meaned variables) but include the yearly averages across the panel units of the dependent and explanatory variables. These average terms are known as Common Correlated Effects ($F_t$, so-called CCEs) and are used to purge estimates from the impact of strong cross-sectional dependence, induced by common unobservable shocks that have sector-specific (local) effects ($\lambda_i \cdot F_t$). These terms, in particular, would remove the effect of "third (unknown) factors" that would cause spurious correlation between dependent and explanatory variables. This could occur especially in the presence of commonalities, i.e., when the sectors have similar knowledge bases but innovate independently of each other.[16] In estimation in column (5), the effect of neighbor innovativeness is still significant, hence excluding the risk that our proxies for knowledge spillovers capture the impact of unmeasurable common factors (see Eberhardt et al., 2013). Note, however, that the elasticity of innovation output to both measures of neighbor innovativeness is smaller, signaling an important source of variation in sector innovation, induced by omitted factors, that we are unable to account for in our benchmark regression.

To corroborate the view that patent texts are a meaningful source to represent the technology content of innovations, we estimate a counter-factual regression in which we assign random values of cosine similarity to each pair of technological classes, while preserving bilateral citation flows as weights in the other measure of neighbor innovativeness (column (6)). Not surprisingly, the impact estimated for the text similarity-based measure of neighbor innovativeness is at odds with all our previous regressions, while the coefficient of the citations-based measure of neighbor innovativeness is moderately larger.

In columns (7) and (8), we address another potential source of concern for our estimates, i.e., that they are influenced by the assumption of (slope) homogeneity in the effect of knowledge spillovers while, thus far, we have confined heterogeneity to sector fixed effects. However, as long as sectors structurally differ in gaining from the innovation of linked units, the effect estimated for neighbor innovativeness may be biased. To exclude this risk, we estimate the model with the mean group estimator, i.e., we run sector-by-sector regressions and take the mean robust to outliers of these coefficients (Bond et al., 2010). This procedure has been applied both to the specification that is equivalent to using time dummies (weak cross-sectional dependence; column (7)) and to the specification using CCE terms (strong cross-sectional dependence; column (8)). In either case, the coefficient of the text similarity-based measure of neighbor innovativeness is much larger than in our benchmark regression (column (1)), while the citation-based measure has a much smaller elasticity. All this reveals that there

[15] The CS-DL specification used is shaped as $y_{it} = b_{0i} + b_1 x_{it} + \sum_{p=0}^{P} b_p \Delta x_{it-p} + \epsilon_{it}$, where $\Delta x_{it}$ is the first difference of the explanatory variable, and $p$ is its lag order. The auxiliary variable is used to purge out the regression from the effect of dynamic adjustment, error serial correlation, etc. In Table A.2, $p$ is set to zero, but similar results emerge using a different lag order.

[16] The estimator used in column (5) corresponds to the cross-sectionally augmented version of the ARDL regression developed by Chudik and Pesaran (2015) with homogeneous (slope) coefficients.

**Table A.2**
Long-run estimates for the effect of neighbor innovativeness: Econometric checks.

| | (1) | (2) | (3) | (4) | (5) | (6) | (7) | (8) | (9) | (10) |
|---|---|---|---|---|---|---|---|---|---|---|
| **Neighbor innovativeness** | | | | | | | | | | |
| Citation flows | 0.107*** | 0.128*** | 0.122*** | 0.074*** | 0.0262*** | 0.165*** | 0.028*** | 0.012*** | 0.075*** | 0.043*** |
| | (0.001) | (0.002) | (0.003) | (0.005) | (0.002) | (0.002) | (0.006) | (0.002) | (0.002) | (0.003) |
| Text similarity | 0.541*** | 0.576*** | 0.518*** | 0.449*** | 0.218*** | −0.115*** | 1.687*** | 1.631*** | 0.254*** | 0.415*** |
| | (0.011) | (0.012) | (0.013) | (0.034) | (0.023) | (0.038) | (0.141) | (0.102) | (0.021) | (0.031) |
| **Adjustment parameter** | | | | | | | | | | |
| | −0.101*** | −0.092*** | −0.093*** | | −0.386*** | −0.063*** | −0.266*** | −0.540*** | | |
| | (0.010) | (0.011) | (0.012) | | (0.022) | (0.009) | (0.024) | (0.028) | | |
| Cross-Sectional Dependence (CSD) | Weak<br>Time dummies | Weak<br>Time dummies | Weak<br>Time dummies | Weak<br>Time dummies | Strong<br>CCE | Weak<br>Time dummies | Weak<br>Time dummies | Strong<br>CCE | Weak<br>Time dummies | Weak<br>Time dummies |
| Patent variable | Forw. Cites | Forw. Cites | Forw. Cites | Forw. Cites | Forw. Cites | Forw. Cites | Forw. Cites | Forw. Cites | Forw. Cites | Forw. Cites |
| Matrix weights: | | | | | | | | | | |
| variable | Text | Text | Text | Text | Text | Random | Text | Text | Text | Text |
| scheme | Direct | Direct | Direct | Direct | Direct | Direct | Direct | Direct | Direct | Direct |
| Model | ARDL (2, 1) | ARDL (4, 3) | ARDL (6, 5) | CS-DL | ARDL (2, 1) | ARDL (2, 1) | ARDL (2, 1) | ARDL (2, 1) | Negative binomial | Inverse hyperbolic sine |
| Parameter | Homogeneous | Homogeneous | Homogeneous | Homogeneous | Homogeneous | Homogeneous | Heterogeneous | Heterogeneous | Homogeneous | Homogeneous |
| Obs. | 5,632 | 5,376 | 5,120 | 5,760 | 5,248 | 5,504 | 5,632 | 5,376 | 4,608 | 5,888 |
| R-squared | 0.141 | 0.146 | 0.152 | 0.612 | 0.663 | 0.147 | 0.089 | | | 0.5402 |
| RMSE | | | | | | | | 0.15 | | |
| Log-likelihood | | | | | | | | | −28,794 | |
| Alpha | | | | | | | | | 0.172 | |

**Notes**: Long-run (dynamic) estimates derived from an ARDL regression are reported in columns (1)-(3) and (5)–(8). These use a different lag structure, as indicated in the table. Long-run (dynamic) estimates derived from a CS-DL regression are reported in column (4). Static estimates derived from a negative binomial and the inverse hyperbolic sine transformation are shown in columns (9) and (10). In all regressions excluding columns (9) and (10), variables are expressed in logs. Heteroskedasticity-Autocorrelation Consistent (HAC) standard errors are in parentheses. All regressions use sector-specific fixed effects. Expect that in columns (5) and (8), all regressions allow for the effect of common time shocks (time dummies) using variables expressed in deviation from their yearly means; this accommodates the effect of weak cross-sectional dependence (CSD). Common Correlated Effects (CCE), computed as cross-sectional means of all variables of the model, are used in columns (5) and (8); this accommodates the effect of strong cross-sectional dependence (CSD). All regressions consider homogeneous slope coefficients excluding estimates in columns (7) and (8), which assume heterogeneous slope parameters. Innovation is measured by the forward cites-adjusted number of patent counts in all regressions. Patents are univocally assigned to one primary CPC class. The matrix of technological proximity is always based on the pairwise cosine similarity of patent texts, with the exception of column (6), where the proximity values are assigned randomly. All estimates use data at the 3-digit level of technology classes (128 sectors). *** , ** , * denotes statistical significance at the 1, 5 and 10% level, respectively.

are large disparities across sectors in exploiting external knowledge sources.

Next, we take into more serious consideration the count data nature of the dependent variable and check whether our estimates reflect the log-linearization of Eq. (2). To address this issue, we run our knowledge production function as a negative binomial regression, using forward cites-adjusted patents as the left-hand side variable and, as regressors, our proxies for neighbor innovativeness, expressed in logs. In this way, the estimated parameters can be interpreted as elasticities and are comparable to our previous results. The negative binomial is estimated as a static regression pooling time-demeaned data among cross sections, where the fixed effect of each sector is approximated by the pre-sample mean values of innovation outcome (Bloom et al., 2013; Igna and Venturini, 2023). Pre-sample fixed effects are computed as the average of the dependent variable over a time window preceding the interval of the regression; they should capture systematic differences existing across sectors in the patent propensity, in connecting to the other innovating units and to their ability to gain from external knowledge. Our count data regression yields lower elasticities for both measures of knowledge spillovers (0.075 and 0.254 respectively, column (9)) that, however, remain overall consistent with the results of our linear regression (column (1)).[17]

Lastly, we account for the bias that may be produced by the linear transformation, $\ln(1 + \Delta N)$, used to handle observations with zeros. Accordingly, we run our benchmark regression using the inverse hyperbolic sine transformation of original variables (not logged) as proposed by Bellemare and Wichman (2020). This is run as a static regression with sector and year fixed effects, obtaining an elasticity of 0.043 for the citation-based measure of neighbor innovativeness and 0.415 for the one based on text similarity (column (10)). This indicates that our results in column (1) are not plagued by log-linearization.

[17] We compute the pre-sample mean of the dependent variable over the period 1976 and 1985 and use the interval from 1986 to 2021 as the regression period.